version: 19 Jan 2026

# Graceful forgetting: memory as a process

Alain de Cheveigné

Centre National de la Recherche Scientifique (CNRS), France; Ecole Normale Supérieure (ENS), 29 rue d'Ulm, Paris, France; University College London (UCL), United Kingdom

## Abstract

A rational framework is proposed to explain how we accommodate unbounded sensory input within bounded memory. According to this framework, memory is stored as a statistic-like representation that is repeatedly summarized and compressed to make room for new input. Summarization of sensory input must be rapid; that of abstract trace might be slower and more deliberative, drawing on elaborative processes some of which might occasionally reach consciousness (as in mind-wandering). Short-term sensory traces are summarized as simple statistics organized into structures such as a time series, graph or dictionary, and longer-term abstract traces as more complex statistic-like structures. Summarization at multiple time scales requires an intensive process of memory curation which might account for the high metabolic consumption of the brain at rest. Summarization may be guided by heuristics to help choose which statistics to apply at each step, so that the trace is useful for a wide range of future needs, the objective being to “represent the past” rather than tune for a specific task. However, the choice of statistics (or of heuristics to guide that choice) is a potential target for learning, possibly over long-term scales of development or evolution. The framework is intended as an aid to make sense of our extensive empirical and theoretical knowledge of memory and bring us closer to understanding it in functional and mechanistic terms.

## I. Introduction

"*I remember him...*" opens *Funes the Memorious* (Borges 1944), the story of a man with a perfect memory. Bedridden after a fall from horseback, his perception and recall were infallible, but Ireneo Funes lacked the ability to generalize or make abstractions: "*it bothered him that the dog at three fourteen (seen from the side) should have the same name as the dog at three fifteen (seen from the front)*." This "reminiscence" of Borges is as much a construct of his imagination as a readout of his own exceptional memory, but who cares? *Si non è vero, è ben trovato*. As for every reminiscence, the past is not present to contradict us.

This paper looks at abstraction, remembering, and forgetting, revisiting some of the questions raised by Borges. How to deal with endlessly accumulating data? Should we, and can we, remember them all, or instead forget some, and if so which, and how? These questions confront our senses, which endlessly absorb sensory input, scientists accumulating astrophysical or neurophysiological data, social media companies harvesting information on their users, societies burdened by their past, and so-on. Among issues are the cost or limits of storage, the time and resources required to search within memory, the usefulness of old memories relative to new, and practicalities of remembering and forgetting.

Memory skills were once greatly valued: techniques to expand mnemonic capacity formed an "art of memory" that was part of the art of Rhetorics (Yates 1966). Printing in the 16th century relaxed the need to carry around vast sums of knowledge in our head, and memory then began to be perceived as antagonistic to mental skill. Arguments made at that time include: *the brain has to be purged of old learning to make room for the new* (Rabelais), *memory and understanding depend on incompatible proclivities* (Huarte), *learning stifles the ability to discover* (Montaigne), *reduction of things to their causes obviates the need for memory* (Descartes). As Montaigne put it, "*better a head well-made than a head well-filled*" (Weinrich 1997). The ability to forget is said to be a factor of creativity (Storm and Patel 2014) and emotional well-being (Nørby, 2015). Nowadays, digital tools offload our memory to the Internet, accelerating the trend initiated by the inventions of writing, printing, and recording (Sparrow et al. 2011; Marsh and Rajaram 2019; Storm and Soares 2024). And yet, data giants such as Google, Facebook, or Amazon thrive on accumulating data, so there must be some value in an extensive memory. Many of us would appreciate a boost in our ability to remember birthdays or memorize poems and music.

Memory has been likened to a retrospective arrow, a "telescope pointed at time" (Proust, 1913; Tulving, 2002), but it is also a prospective tool to manage the future (Schacter et al., 2012; Nobre and Stokes, 2019). Guessing the next state of the world, or planning the best actions to influence it, or to extract more knowledge from it, are of obvious survival value, and for this we rely on knowledge gleaned from the past.

In this paper I outline a rational framework for memory based on an incessant process of statistical summarization, sketch links with what is known of human memory, and explore implications for machine learning. The framework is characterized by its reliance on statistics as a "lingua franca" for memory, and by the assumption that the entire contents of memory, old and new alike, are incessantly revised to make them ever more concise and compact. An apt metaphor is *rumination*, the process by which certain animals, e.g., a cow, regurgitate their fodder, chew it some more, and swallow it again[1].

[1] This metaphor was used also by Gershman et al (2017), and much earlier by Augustine (397): "*memory is like a stomach for the mind*" (Book 10, ch. 14). It is distinct from the term "rumination" used in psychiatry.

## II. A rational framework

The phenomenology of memory is complex, as reflected by the many terms employed (e.g., *remember*, *know*, *learn*, *train*, *memorize, recollect*, *recall*, *retrieve*, *reminisce*, *recognize*, *remind*, *recode*, *forget*, *memory*, *trace*, *engram*, *retention, knowledge*, *meme*, *remembrance*, *memoria*, *condensation*, *redintegration*, *familiarity*, *oblivion*, etc.), and their various patterns of semantic overlap and polysemy. The memory literature is broad, experimental approaches are many, and it is hard to build a synthetic understanding. Rational analysis helps by identifying mechanisms that must be implemented to get a job done, so that empirical results can be reviewed in terms of how those mechanisms are implemented in the brain. Examples of the rational approach to memory are Anderson and Milson (1989), Chater and Oaksford (1999), Gershman (2014), Griffiths et al (2015), Sims (2016), Lieder and Griffiths (2020), Gershman (2021), Bates and Jacobs (2020), Cox and Shiffrin (2024). Data science and machine learning have lately come to play a similar role (e.g., Richards et al 2019; Lindsay 2020; Rae 2021; Mattar et al 2022).

What is memory? A broad definition is "*a message that we send to our future self*" (or "*that we receive from our past self*"). This definition hints at the need for resources to encode such a message and decode and attend to it, the medium that carries the message, the contents of that message, and the real-world referent of the message, each of which is sometimes referred to as "memory" (an example of polysemy that plagues the field according to Tulving 1972 and Cowan 2017; Roedinger and Uner 2024). It suggests also an analogy with a *communication channel*, so we can draw on rate-distortion theory to formalize the effects of memory capacity limits (akin to channel capacity) or decay (akin to noise) (Sims 2016; Bates and Jacobs 2020; Nagy et al 2020; Gershman 2024).

However, the analogy with a communication channel is strained. There is no point at which a message is coded: rather, the stream of new information about the world is squeezed into an existing message, which therefore must be *recoded* to make room. If I am allowed only 1000 words to keep an up-to-date diary, I must necessarily rewrite it every day. This could take various crude forms, such as deleting the old, or ignoring the new, but intuition tells us that some combination of old and new might do better justice to both. Implementing that combination requires an ongoing process that I will argue is *necessary* (for a competitive edge in life's game), *costly* (for the organism to implement and maintain), and *complex* (for us to understand).

Why remember? The distal goal is obviously survival of the individual, group, society, or species. More proximal goals are to predict the future stimulus ("predictive coding") or the future state of the world (Ha and Schmidhuber 2018), or to plan the next move (Tishby and Polani 2011, Kroes and Fernández 2012; Salge et al 2014; Mattar and Lengyel 2022). Arguably, the latter goal is what really counts, but the other two are useful subgoals. An additional role of memory, that will become clear later on, is to help manage memory itself.

What to remember? Arguably, *everything*: every bit of information might turn out to be useful, if only to later discover better compression rules remove redundancy (Schmidhuber 2009). Compression and pruning have other benefits, for example to achieve abstraction, or to expedite search, or to get a head start on future processing. However, *space permitting*, there is no reason to prune now rather than later: we could store both the compressed record and the original and reap the benefits of both. The only serious reason to discard the original is limited storage space; an organism gifted with unlimited storage (and the means to keep it organized) might well want to remember everything.

## For or against boundless memory

*Does* the human brain keep every bit of sensory input? Memory scholars are neatly divided on this hypothesis, some finding it quite plausible, the others quite implausible. Our everyday experience of forgetting argues against it, however that could reflect loss of conscious *access* to memory traces, rather than loss of the traces themselves (Davis and Zhong 2017). Indeed, reports of individuals with highly detailed memory, on the model of Funes, could be seen as an existence proof for boundless memory. Studies that show massive implicit memory for sensory patterns (Agus et al 2010; Utochkin and Wolfe 2018; Bianco et al 2020; Bastug et al 2026) seem to have a similar implication. The debate could be illuminated by comparing brain capacity with cumulated sensory information rates, but both are hard to define and estimate, and results of such calculations are very disparate (Hooke 1682; Burnham 1888; Jacobson 1951; Barlow 1959; Standing 1973;

Landauer 1986; Dudai 1997; Cunningham et al 2015; Jenkins et al 2018). A rough estimate of capacity can be derived from counting neurons and synapses (roughly $10^{11}$ and $10^{14}$ respectively), but other candidate substrates such as synapse growth and rewiring, dendritic compartments, astrocytes, or DNA or RNA (Poirazi et al 2001; Chklovskii et al 2004; Knoblauch et al 2014; Languille and Gallistel 2020; Kastellakis et al 2023; Kozachkov et al 2025), have potentially a much larger capacity. The boundless memory hypothesis, while costly, is not easy to dismiss on the grounds of limited capacity of a neural substrate.

A second argument has to do with *search*: assuming that everything is retained, how do we find a pattern of interest? Funes "*knew by heart the forms of the southern clouds at dawn on the 30th of April 1882*", and could compare them in his memory with "*the mottled streaks on a book in Spanish binding he had only seen once*", and with "*the outlines of the foam raised by an oar in the Rio Negro the night before the Quebracho uprising*". This feat requires, in addition to a massive memory, the ability to perform efficient search within that memory so as to discover comparable patterns. Brute force search for pairs of matching elements within a store of size $N$ has a computational cost $O(N^2)$, infeasible for large $N$ without an index, and finding three-way matches is likely harder still. Popular wisdom argues for discarding things merely to reduce clutter (Kondo 2014; Vitale 2018), and the case for forgetting for that reason has been made repeatedly in the memory literature (e.g., Weinrich 1997; Schacter 1999; Connerton 2008; Richards and Franklin 2017; Davis and Zhong 2017; Fawcett and Hulbert 2020).

A third argument has to do with an uneven distribution of relevance over time. Suppose that I care about both the past and the present: for simplicity let's say half my attention is devoted to today and the other half to yesterday and previous days (Fig. 1). Tomorrow, each of those halves will get *one quarter* of my attention, the next day *one eighth*, and so on, suggesting an exponential fading of relevance with time. Such a trend has been observed empirically for books in a library (Burrell 1985), web pages (Jovani and Fortuna, 2007), and citation patterns in the scientific literature (Golosowsky and Solomon 2016; Pan et al 2018; Candia et al 2019). Competition from new input is compounded by the fact that older information might be less applicable to a newer world. Relevance is more thinly distributed over the remote than near past.

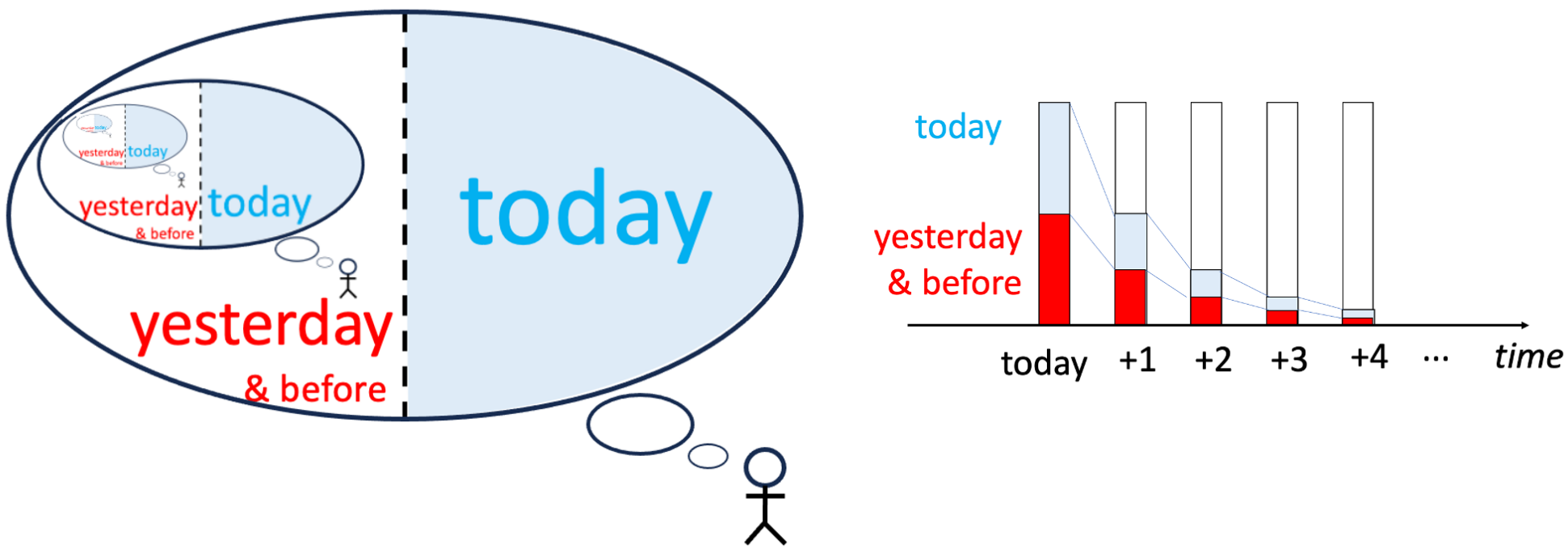

Fig. 1. Relevance tends to decrease with time due to competition from new input. If attention is divided equally between today and previous days, tomorrow each part will capture one quarter of my attention, the day after one eighth, and so on. A similar, exponential trend is observed for library books, web pages, and scholarly citations: the old competes with the new (and loses).

To summarize: three arguments against a boundless memory are *storage limits*, *search cost*, and *uneven relevance*.

The alternative to keeping all the information is to discard some, but this opens a ream of new issues. The stream of observations is relentless, and so discarding must occur repeatedly, requiring an ongoing *process*. The representation (or "data type") must permit repeated down-sizing while still retaining value for future use, which motivates assimilating memory to *statistics*, as elaborated further on. Statistics are of variegated sorts, and repeatedly the decision must be made as to which statistic to choose, and with which parameters. Bits discarded are irremediably lost, so the decision must be wise. In a nutshell: optimal use of past information, given limited storage, requires a specific form of representation, an ongoing process of abstraction, and a smart decision process to guide it.

## Statistics

To qualify as "memory," a trace must be informative about the original observations, and as discussed it must also allow repeated summarization. *Statistics* have these properties, at least when applied to numbers, and, in this paper, I will use them as a model for memory. There is necessarily a gap between the model and the real thing, but I'll discuss various ways in which it can be narrowed, if not bridged. These include enhancing the representation by assembling simple statistics into complex structures, augmenting them with a *dictionary* structure that can represent arbitrary patterns or events, and stretching the definition of "statistic" to capture abstract entities of a possibly non-numeric nature (Fig. 2). The gap is narrowed also by noting that statistics have been widely invoked as perceptual representations in vision and audition (e.g., Ariely 2001; McDermott et al. 2013), and that numerical embeddings have had success in modeling symbolic or conceptual entities (Landauer and Dumais 1997; Kumar 2021).

By itself, a sample of a statistic such as *mean*, *covariance*, or *cardinality* has limited expressive power, but statistics can be aggregated to capture a more complex distribution. Three basic structures are of particular interest: *time series*, *dictionary*, and *hierarchical index*.

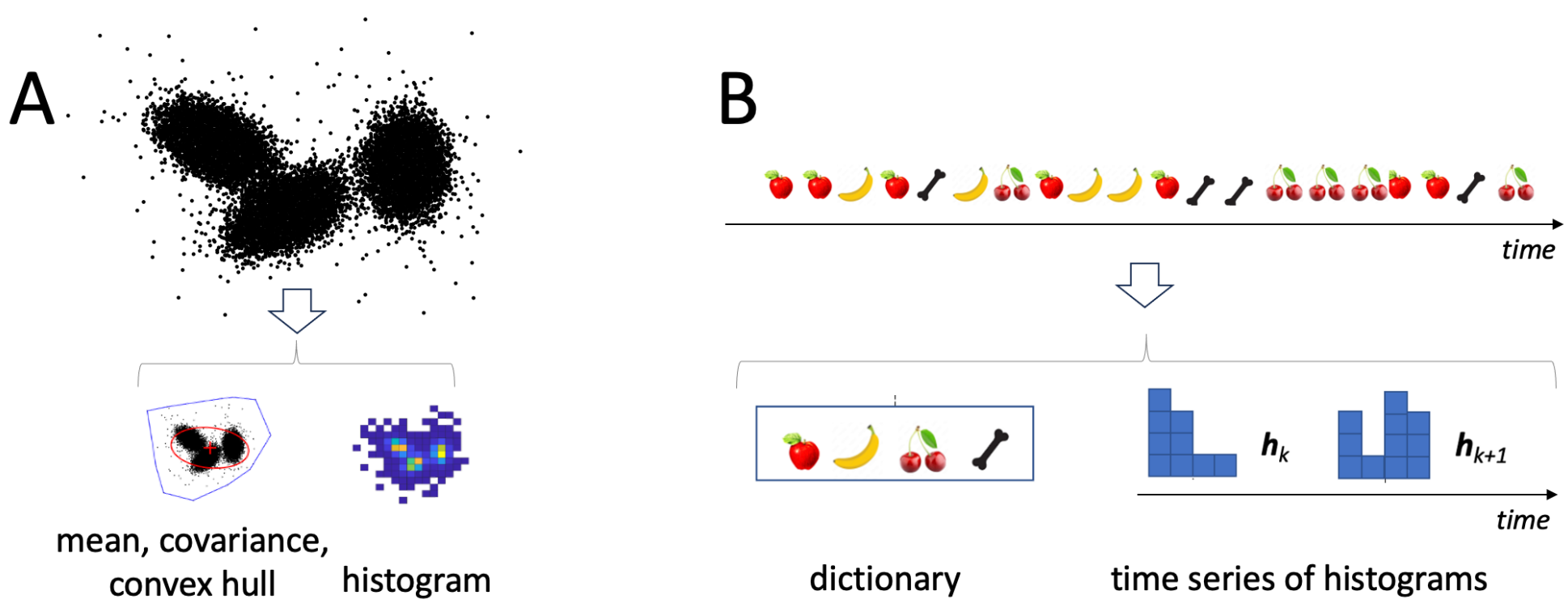

Fig. 2. Examples of simple statistics. A: A lumpy cloud of 2D points (top) can be summarized by its mean, covariance and convex hull (bottom left) or as a histogram (bottom right). B: A time series of non-numerical events (top) can be summarized by associating a dictionary (bottom left) with a histogram that counts occurrences of each entry. The histogram itself can be instantiated as a time series (bottom right).

The time series inherits the temporal order of incoming observations (although the sampling might be non-uniform and the samples diverse, e.g., each a different type of statistic). The dictionary is defined as an unordered collection of entries, which can be associated with bins to form a *histogram* that allows occurrences of these entries to be tallied (Fig. 2B). An entry might consist of an interval of values, as in a classic histogram, a prototype of some category of interest (possibly non-numerical), or a statistical structure that parametrizes a distribution characteristic of a category. The *hierarchical index* is a tree with statistics at its nodes for the purpose of search by value. The time series captures the temporal orientation of the world, the dictionary its lumpiness, and the index helps find items based on their value. Additional statistical structures might assist search, for example various forms of graph such as a semantic network (Steyvers and Tenenbaum, 2005; Kemp and Tenenbaum 2009; De Souza et al 2021).

As mentioned, a useful property of the representation is the ability to summarize it repeatedly. Consecutive samples of a time series might be aggregated and replaced by a single sample, entries of a dictionary might be merged and the corresponding bins summed, and so-on. To make this work smoothly, it is useful for the basic statistics to have a property that I call *scalability*: the value that summarizes a set of observations can be derived from values that summarize disjoint subsets. This may require some care: whereas cardinality is scalable, the mean is scalable only in association with cardinality, and variance only in association with mean and cardinality. The appeal of a scalable statistic is that its value is the same whether it is derived directly from observations, or via multiple rescaling steps.

Yet more complex structures can be assembled to achieve a complex, fine-grained statistical representation over a long term. A dictionary might form one sample of a time series, or an interval of a time series might serve as an entry of a dictionary. More complex structures are potentially more effective (better fit to the empirical distribution of observations for a given storage cost), but possibly harder for an organism to implement, and for us to conceptualize.

In sum, statistics parametrize the *empirical distribution* of the data that they summarize. From a different perspective, they describe the *texture* of the original data. Alternatively, the statistical representation is an *abstraction* that captures a *gist* or *trend* possibly useful for the future. From yet another perspective, it is the result of *attention* to particular aspects of the past and neglect of the rest, or a form of *dimensionality reduction* which may assist pattern recognition if the dimensions retained are wisely chosen. From yet another perspective, it is akin to *lossy compression* of the input, or *coding* by a lossy channel (Bates and Jacobs 2020), or a layer of a *neural network*. These various interpretations are all consistent with the idea of "making the most" of data before they are lost.

## How can a statistical representation be used?

A statistic can parametrize a distribution to estimate the *probability* that a value or event, now forgotten, was observed. We can judge whether a new observation is "surprising," or search an index for a node consistent with the observation, or drive a generative process to produce samples typical of previous observations, for example to implement Bayes formula for the purpose of inference (Orban et al. 2016), or to compare traces based on a statistical measure as elaborated below. In this perspective, the statistic serves to represent (or approximate) the values of observations. A different perspective uses the statistic to represent a *pattern* of samples within a sensory stream. For example, the short-term power spectrum is an abstract representation of the spectral content that factors out irrelevant phase differences. Repeated summarization creates a hierarchy of increasingly abstract and concise representations, high-resolution statistics (at the leaves) representing recent, concrete traces, and lower-resolution statistics (near the root) representing long-term abstract memories.

A difficult issue must be addressed. Two exemplars of the same pattern (or indeed, the exact same sensory data) might be summarized by different statistics, for example on one side by mean and covariance, and on the other by a convex hull, or a histogram. Or the statistics might be the same but sampled according to a different schedule. If we are to make useful comparisons across memory (as when Funes compared clouds, foam, and patterns on a book cover), we need to be able to recognize that representations are similar despite the different parametrizations. Comparing the traces themselves is useless. Instead, we must resort to a statistical measure such as the Kullback-Liebler (KL) divergence, defined as

$$D_{KL}(\mathcal{A} \parallel \mathcal{B}) = \sum_x \mathcal{A}(x) log(\mathcal{A}(x)/\mathcal{B}(x)) \qquad (1)$$

where $\mathcal{A}$ and $\mathcal{B}$ are distributions over a feature space indexed by $x$. Its value is close to zero if $\mathcal{A}(x)$ and $\mathcal{B}(x)$ are everywhere similar, or if $\mathcal{A}(x)$ is small wherever they do differ (i.e., they rarely differ) (Fig. 3). In order to apply Eq. 1, a probabilistic model must be known or chosen (e.g., Gaussian, or uniform) so that the statistics (e.g., mean $m$, variance $\sigma$, etc.) can be translated to a full-blown distribution (e.g., $\mathcal{A}(x) = \mathcal{N}_{m,\sigma}(x)$).

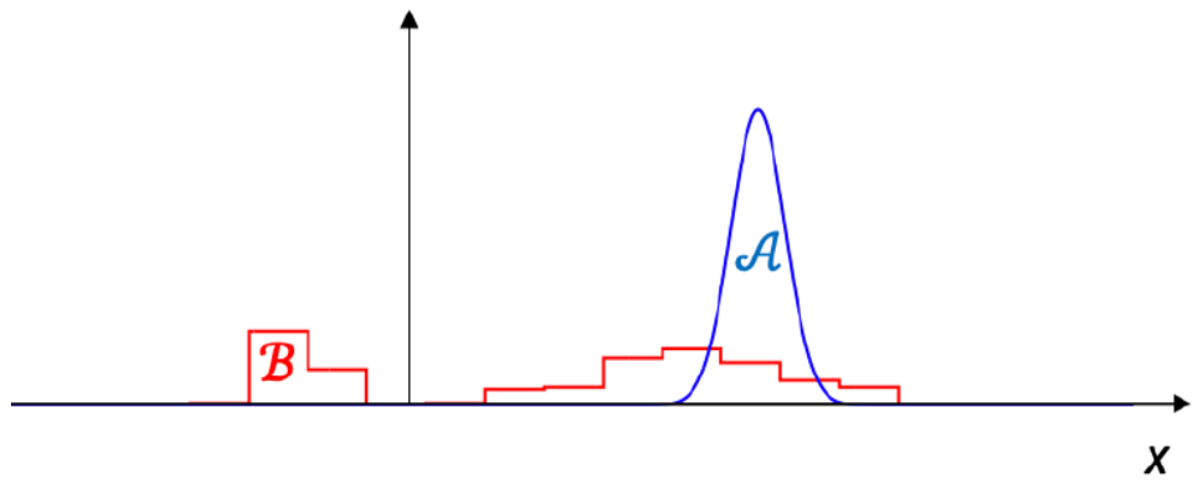

Fig. 3. Two distributions, each with a different parametric model (Gaussian for $\mathcal{A}$ parametrized by mean and variance, piecewise constant for $\mathcal{B}$ parametrized by a histogram). Kullback-Liebler (KL) divergence allows such diverse distributions to be compared. In this example, $D_{KL}(\mathcal{B} \parallel \mathcal{A})$ is almost infinite because $\mathcal{B}(x)$ has large values over a range where $\mathcal{A}(x)$ is tiny, indicating that it is unlikely that observations summarized by $\mathcal{B}$ all come from the process that produced the observations summarized by $\mathcal{A}$. The reverse is less unlikely: samples drawn from $\mathcal{A}$ might conceivably fit $\mathcal{B}$.

Interestingly, KL divergence is asymmetric, so the questions "*is new pattern $\mathcal{A}$ consistent with my memory of previous pattern $\mathcal{B}$?*" and "*is my memory of previous pattern $\mathcal{B}$ consistent with new pattern $\mathcal{A}$?*" might allow for different answers. It is tempting to "symmetrize" the measure (e.g., by taking the average of $D_{KL}(\mathcal{A} \parallel \mathcal{B})$ and $D_{KL}(\mathcal{B} \parallel \mathcal{A})$) to make it more like a distance, but that might blind us to a possibly genuine asymmetry. The present might indeed be surprising in view of the past but not the past in view of the present (or vice-versa). Applying KL divergence to statistical traces is more complex than applying Euclidean distance to points within a feature space, but this complexity is inescapable. If we adopt the memory-as-statistics hypothesis, we must accept that the brain approximates a statistical measure of similarity to allow traces to be compared.

## The process

At the same time as it preserves previous information, memory must accept new input. This requires reorganization, implying a *process*. The process transforms the memory trace at time $t$, together with input at time $t$, into an updated trace at time $t + 1$:

$$\mathcal{O}_t, \mathcal{M}_t \rightarrow \mathcal{M}_{t+1} \qquad (2)$$

where $\mathcal{O}$ represents observations and $\mathcal{M}$ memory.

Assimilating the trace to a statistical structure as described earlier, every node of the new structure could derive from some combination of all nodes of the old but, again, some simplifying assumptions are useful to facilitate understanding. If memory is structured as a *time series*, each sample of the new series can be derived from a *segment* of the old. This conserves temporal proximity and order, but begs the questions of how to define *segment boundaries* (the segmentation, or change-point detection problem of data science, Aggarwal 2007), and which statistic(s) to use to merge the samples. Any rule might work in principle (e.g., "replace each pair by their mean") but more complex rules might result in a better representation, as discussed shortly. If memory is structured as a *dictionary*, it can be made more concise by merging entries, or by making individual entries more concise, or by deleting unneeded entries. If it is structured as a *hierarchical index* (tree), it can be made more concise by pruning the leaves or twigs.

**An illustrative example.** Let us suppose a multivariate stream (time series) of observations (Fig. 4, top) that is accumulated into a series of "buffers", the first containing a series of raw samples, and the following statistics of increasing concision. For example, a sample might consist of the *cardinality*, *mean* and *covariance* over a group of observations. As time proceeds, consecutive samples of each buffer are summarized into samples of the next buffer. Together, these buffers represent past values of the stream with a resolution that decreases exponentially with age. With appropriate scaling rules, this "memory" might accommodate unbounded input within finite storage.

In this example, cardinality, mean, and covariance are unequally expensive (1, $d$, and $d(d-1)/2$ numbers per sample respectively where $d$ is the number of channels) and therefore it might be advantageous to sample them at different rates. Moreover, if the observations are non-stationary, busy intervals should be sampled more densely than quiescent intervals, leading to a non-uniform time series of statistics (Fig. 4, bottom).

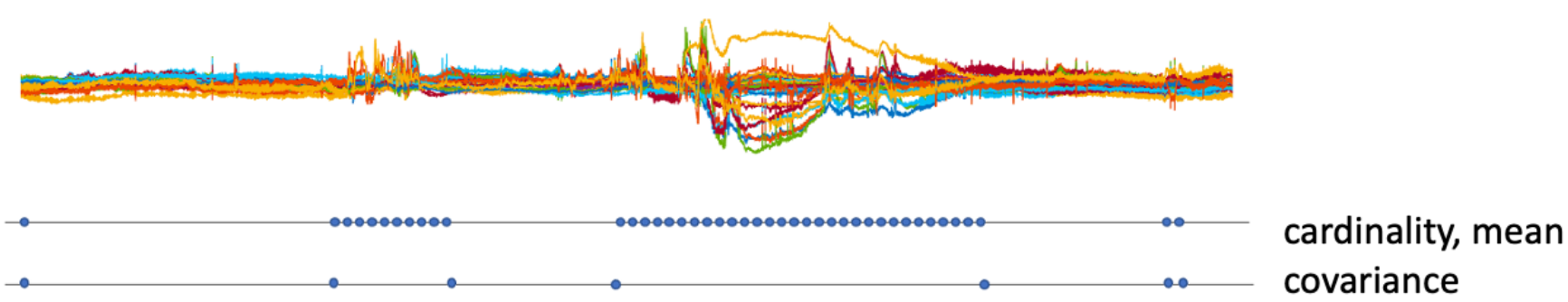

Fig. 4. Multivariate time series (top) summarized by time series of cardinality, mean, variance and covariance (each dot is a sample of the statistic). Sampling density can vary according to the time-varying properties of the signal, and expensive statistics can be instantiated with a coarser resolution.

Continuing with this example, a segment of observations representing a *pattern*, *event*, or *episode* can be stored as an entry in a dictionary. Subsequent occurrences of that pattern are then coded as *pointers* to the appropriate entry (Norris 2017), and those pointers subsequently grouped and their count coded within one bin of a histogram. The histogram itself can be instantiated repeatedly to form a time series of histograms, samples of which can be later merged over time. The dictionary entries can be condensed, merged, or deleted as needed, and so on. Rather than deleting samples after summarization, intermediate results can be conserved to populate layers of a hierarchical search index. Summarization ceases when the need to free storage eases, leaving a compressed yet detailed parametrization of the distribution of the original data, more detailed over the recent than remote past. This example is not intended to resemble an actual memory trace, but rather to illustrate concretely curation operations that must occur in memory.

## Heuristics

Statistics offer a flexible representational model, but with flexibility comes a new problem. At each summarization step, a decision must be made as to which records to delete or replace by a summary, which statistic to choose (e.g., mean/covariance vs convex hull), and with which parameters (sampling resolution, etc.). Those decisions affect how well the memory will meet future demands. Those are unknown at the time of the decision; the best that we can do is estimate a *need probability* (Anderson and Milson 1989; Griffiths et al 2015), on the basis of information at hand when the decisions are made. A number of *heuristics* may be useful in this context:

1. *Recency*. This heuristic is automatically enforced by repeated summarization, as older content will have been summarized more times than new.
2. *Novelty* (Duszkiewicz et al 2019), *KL divergence*, s*urprisal*, *prediction error* (Gershman et al 2017). A node with a unique statistical distribution should be kept, predictable observations can be discarded.

3. *Recurrence* (a pattern that recurs is unlikely to be random), *slow variation*, as in Slow Feature Analysis (Wiskott and Sejnowski 2002). Physical objects have inertia, causal effects may take time to unfold, etc.
4. *Independence*, *decorrelation, discrimination*. Redundant dimensions can be discarded (e.g., PCA, ICA) (Attneave 1954), discriminative dimensions can be prioritized (e.g., LDA).
5. *Prior access* (Anderson and Schooler 1991, Awh et al 2012; Sekeres_2016): attended once, interesting ever. *Reference from other records*: this heuristic is used by the PageRank algorithm to prioritize web pages (Brin and Page 1998; Griffiths et al 2007).
6. *Temporal or causal association* with past *action* or *decision* (Murty et al 2019). It is useful to remember the events that led up to a decision or action, and what happened next.
7. *Temporal or causal association* with past *reward* (Awh 2012; Braun et al 2019), *reward change* (Rouhani et al 2020), *punishment*, *stress*, *emotional state* or *social feedback* (Dunsmoor et al 2015, 2022).
8. *Conscious choice*. Interest, curiosity, directed forgetting (Hertwig and Engel 2016; Fawcett and Hulbert 2020), involvement in "work in progress" and unmet goals (Conway 2008; Cowan et al 2024).
9. *Reliability*, *confidence*, *source monitoring, predictive power* (Kim et al 2016), *depth of processing* (Craik and Lockard 1972; Utochkin and Wolfe 2018; Lin 2024).
10. *Animacy, ancestral priorities* learned via evolution and hard-wired (Nairne et al 2017), *foraging-related information:* keep track of what was already explored/exploited (Fougnie et al 2015).

Heuristics are analogous to the "reduction principles" of Attneave (1954). A heuristic by itself does not trigger deletion, rather it helps *prioritize* which dimensions to delete. Heuristics can be grouped according to the information that they exploit: *regularities among the observations* (heuristics #2-4), *usage patterns* (#5), *data relevance measures* (#6-10). For example, heuristic #2 suggests that if the KL divergences of a first node relative to a second, and of the second relative to the first, are *both* small, the nodes are not worth keeping separate. The forward direction (new relative to old) corresponds to *surprisal* (Baldi and Itti 2010, Gershman et al 2017), however both directions are potentially relevant. Taking heuristic #4, joint diagonalization of the covariance matrices attached to two nodes may reveal that they differ only on one or a few dimensions (as in linear discriminant analysis).

Choosing one *heuristic* over another prioritizes different aspects of the information, thus affecting future usefulness. This *second level of decision* in the memory process could be a target for learning (or meta-learning) over a long term, including developmental and evolutionary scales. There is a tension between optimizing based on *known needs* versus acquiring general-purpose knowledge for *unknown needs*, reminiscent of the classic exploit/explore divide. Ultimately choices must be tuned based on “what worked” in the past, possibly remote or ancestral.

This concludes the “rational analysis” part of this paper. Based upon it, we know that memory must involve an incessant process of summarization at every time scale and level of abstraction. Memory requires a representation that supports repeated summarization, for which statistics are a convenient model. The many degrees of freedom imply a decision process (choice of statistics) that may be guided by heuristics that themselves require prioritization. Comparison between memory traces requires a non-trivial operation akin to KL divergence. The next section reviews known aspects of biological memory in the light of this framework.

## III. Human memory as a process

The purpose of this section is to establish links with the literature on human memory. The story is simple: the rational constraints mentioned earlier *must* be satisfied; identifying *how* they are satisfied gives insights as to how memory works. The topics invoked are important, but the list is long and the busy reader might wish to skip to the Discussion and come back for more detail. For easy reference, the framework is dubbed "memory as a process", MAP.

### Parcellations of memory

Memory has been parcellated in various fashions according to nature, role or time scale. *Primary* memory has been contrasted with *secondary* (James 1890), *procedural* with *propositional* (Tulving 1984), *declarative* with *non-declarative* (Squire et al 1993), *episodic* with *semantic* (Tulving 1972; Conway 2009; Renoult et al 2019; Renoult and Rugg 2020), *long-term* (LTM) with *short-term* (STM) (Shevlin 2020), and further subdivisions have been proposed such as *phonological loop* (Baddeley and Hitch 2019), *conceptual short-term memory* (CSTM) (Potter 1993), and others (Radavansky et al 2022). The MAP framework assumes a continuum from concrete to abstract, but implementation of storage and summarization at different time scales might require specialized modules. The debate between "unitary" and "multistore" memory (e.g., Norris 2017, 2019; Cowan 2019) recalls that between "lumpers" and "splitters" in Darwinian classification (Darwin 1857; Endersby 2009; Vives et al 2023). MAP might appeal to both.

A useful distinction is between *archival* and *computational* roles, the former referring to a remanent trace of the past, and the latter to a temporary "scratchpad" or "working memory" (WM) (D'Esposito and Postle 2015; Oberauer 2019). The MAP framework addresses the archival role. The term "archival" evokes the metaphor of a storehouse of past observations (Hooke 1682; Hintzman 2003; De Brigard 2014), but for MAP the store contains mostly abstractions derived from those observations rather than the observations themselves. It is archival in that it contains all that the organism knows of its past. STM and WM are sometimes equated, presumably because a transient sensory buffer might plausibly serve as a scratchpad for transient mental computations. Results of a computation might be worth archiving in LTM together with, or in lieu of, its antecedents, as suggested early on by Hooke (1682) and Augustine (397), and it has been proposed that STM/WM consists of portions of LTM "activated" to serve the needs of computation (D'Esposito and Postle 2015; Cowan 2017; Kumle et al 2025).

## Forgetting

Forgetting is a matter of regret because every bit of information is potentially useful (Schmidhuber 2009), yet inevitable given the need to accommodate new information. The best we can hope for is to forget in a "graceful" fashion, hence the title of this paper. Loss of information is motivated primarily by the need to reclaim space, but there may be other benefits such as removing over-detailed or stale information (Hollingworth 1910; Schlesinger 1970; Weinrich 1997; Schacter 1999; Wixted 2005; Connerton 2008; Hardt et al 2013, Storm and Patel 2014, Nørby 2015, Hertwig and Engel 2016; Richards and Frankland 2017; Fawcett and Hulbert 2020, Nagy et al 2020).

Forgetting can result from *deletion* (Zhang and Luck 2009), *overwriting* (Lewandowsky and Oberauer 2009), *interference* (Wixted 2004; Robertson 2018), *decay* (Hard et al 2013), or *noise* (Estes 1997, Brady et al 2024), all of which are irreversible processes that lead to *loss of availability*. In addition to these, forgetting might result from *loss of accessibility* which is potentially reversible (Ryan and Frankland 2022). A nominally deleted trace might linger (as when computer memory is freed, or disk blocks unlinked), and thus a presumed-erased trace might in some conditions remain accessible (Frankland et al 2019), reminiscent of a *palimpsest* in which an old inscription can be guessed behind the new (Weinrich 1997; Matthey et al 2015).

Interference from new input might arise in multiple ways. First, new input might displace older traces for lack of space. Second, assigning a new observation to a dictionary entry might update that entry and distort the "memory" that it represented previously (Bein et al 2023). If this leads to a wider statistical distribution, the trace might later be merged with another trace, or pruned (Umberto Eco suggested that adding trivia to a memory is the best way to destroy it, Eco 1988). Third, retrieval of a trace might prioritize it (heuristic #5) at the expense of other traces (Beckinstein 2018). Fourth, redintegration of an abstract trace into a concrete "recollection" might recruit more recent traces (Potter 1993, 2012; Schacter and Addis 2007; Addis 2018), and this could again lead to interference from newer input.

For MAP, forgetting is primarily a loss of *availability*, mitigated by statistical summarization. It has less to say about loss of *accessibility*, which may also contribute to forgetting.

## Statistics

Assimilating memory to statistics is a bold assumption, and one might question whether statistics can represent traces of a qualitative or symbolic nature. However, the success of latent semantic analysis (Landauer and Dumais 1997) and more recently large language models (LLM) suggests considerable expressiveness. Moreover, even for a seemingly non-numerical event, in the long run what counts may be the *number* of times it occurred. Indeed, Barlow (1990) has argued that, in order to detect remarkable patterns of coincidence between events, *all* events must be counted, remarkable or not. This entails a lot of counting.

Statistics have been invoked to represent sensory or perceptual features in vision (Attneave 1954; Ariely 2001; Whitney and Leib 2018) and audition (McDermott et al. 2013; Nelken and de Cheveigné 2013), including low-level features such as hue, size, or orientation (Whitney and Leib 2018), and higher-level features such as emotionality of faces (Haberman and Whitney 2009), gender (Haberman and Whitney 2007), or lifelikeness (Leib et al 2016). Statistics can be gathered in parallel for different levels of abstraction, e.g., color and expression (Haberman et al 2015), or different parts of an image (Sun et al 2018), or streams in an auditory scene (Hicks and McDermott 2024), or strands in an individual's life (Kubovy 2015), and observations may be integrated over disjoint intervals (Waskom and Kiani 2018). There is evidence that the statistical representation is coarser at longer time scales (McWalter and McDermott 2018; Hicks and McDermott 2024) as consistent with progressive summarization. Examples of statistics invoked include *cardinality* (Underwood 1969), *mean* (Haberman and Whitney 2009), *variance* (Michael et al 2014; Khayat and Hochstein 2018), *range* (Lau and Brady 2018; Khayat and Hochstein 2018), as well as more complex statistical representations akin to a *histogram* (Chetverikov et al 2017a; Khostov et al 2025).

As mentioned earlier, the representational power of statistics is boosted by assembling them into structures. A *time series* can model how incoming sensory stream is buffered as required by e.g., echoic or iconic memory (Coltheart 1980; Scharnowski et al 2007; Rensink 2014). Over a longer timescale it might capture the historical axis of experience, or the temporal structure of specific events or episodes as required for episodic memory (Conway 2009; Farrell 2012; Rubin and Umanath 2015). A *dictionary* might serve as a model of a semantic or reference memory (McClelland et al 1995; Eichenbaum 2017; De Souza et al 2021), as well as supporting the recording of arbitrary events. An array of counters associated with a dictionary might tally event occurrences as a *histogram* to count events of each kind (Underwood 1969; Barlow 1990). A *hierarchical index* caters for efficient search, and other graph structures might support higher-level abstractions such as "schemas," "scripts," "mnemonic structures" or "knowledge" (Kemp and Tenenbaum 2008; Ghosh and Gilboa 2014; DeSouza 2021; Shearer et al 2025). Statistics offer a model that allows the requirements of boundless summarization to be discussed in concrete terms.

## Recognition and recall

Access to memory is thought to take two forms known as *recognition* and *recall* (Raaijmakers and Shiffrin 1992; Umanath and Coane 2020). The former (also referred to as *familiarity* or *knowing*) involves scanning memory to check for presence of a given item, for example to determine whether a face has been seen before. The latter (recall, also referred to as *recollection, retrieval* or *reminiscence*) additionally returns information such as the name associated with the face or details of an encounter with that person. We might assume naively that what is recalled matches what was stored initially, diminished by what was forgotten (the "storehouse" metaphor), and much of memory science has been devoted to determining the "laws of forgetting" using lists of words, in the tradition of Ebbinghaus (1885/1913).

However, it was recognized early on that recall is a more complex process. For Augustine (372, book 10, ch11), memories are "re-collected" from a scattered state (in his words: "*cogo [draw together] and cogito [cogitate] are related*"). It has since been understood that the process is not just *recollective* or *reconstructive*, but also *creative*: recall typically includes details completely different from what was observed (Bartlett 1932; Loftus and Loftus 1980; Tulving 1985a; Potter 1993; Schacter and Addis 2007; Gallo 2010; Patihis et al 2013; De Brigard 2014a; Lewandowsky and Oberauer 2015; Fernandez 2015; Addis 2018; Barry and Macguire 2019; Mildner and Tamir 2019; McWalter and McDermott 2023; Spens and Burgess 2024).

Recollection produces a concrete trace from an abstract representation for the benefit of our conscious selves (e.g. to think about an event), or to communicate our memory to others (de Brigard, 2014a; Mahr and Csibra 2018; Corballis 2019). We might not doubt that the result is veridical, for lack of contrary evidence, and an inexact recollection might even serve better our purpose ("*si non è vero*..."). A similar generative process has been invoked in *future planning* (Kahneman and Miller 1986; Moscovich 2008; Hassabis and Maguire 2009; Schacter et al 2012; Clark 2013; Barry and Maguire 2019; Mildner and Tamir 2019; Butz et al 2021; Zacks et al 2022), *creativity* (Benedek et al 2023), *mind wandering* (Christoff et al 2016; Mildner and Tamir 2019) and *dreaming* (Domhoff 2011; Llewellyn 2013). Generation is also inherent in *analysis-by-synthesis* models of perception (Helmholtz 1867; Yuille and Kersten 2006)

According to the memory-as-statistics viewpoint, the gist-like statistical trace represents a distribution that can be evaluated at the value of a probe (recognition) or sampled to produce a concrete trace (recall). The recalled trace is unlikely to match any earlier observation, even though, by construction, it is consistent with our *memory* of that observation. Due to the mismatch between recalled and original observations, memory may appear noisy (Brady et al 2024) even if the representation itself is noiseless. The distribution might be modified by the query (the "synergistic ecphory" of Tulving 1984), and the fact that the generative sampling process must draw on more concrete representations might lead to bias from more recent observations (Brady and Alvarez 2011), motivating the claim that "the external environment is an integral part of the memory system" (Craik 2021, p60).

## Encoding

Input to memory is, potentially, everything worth remembering: sensory input, decisions, rewards, emotions, thoughts, and even meta-memory (memory of remembering or of forgetting, Augustine 397). The input is pruned by attentional processes, but if we assimilate those to a form of summarization, immediately applied, input to memory is *everything* observable. Attention is applicable alike to sensory input or to memory (Chun et al 2011, Lindsay 2020; Olivers and Roelfsema 2020), and attentional decisions themselves depend on memory (Woodman et al 2013; Nobre and Stokes, 2019). Sensation, perception, and memory are intimately associated (Demany and Semal 2007; Nelken and de Cheveigné 2013; Christophel et al 2017).

A high-rate sensory stream must be processed rapidly (the "now or never bottleneck" of Christiansen and Chater 2016), and therefore early stages are likely automatic. Sensory transduction operates an initial layer of summarization in the sense that peripheral filtering (in the cochlea or retina) entails convolution with a temporal and/or spatial kernel, which results in a statistic (weighted sum). Mid-level features such as the *modulation spectrum* (Dau et al., 1997; McWalter and Dau, 2017), *second-order modulation spectrum* (Lorenzi et al 2001) and between-channel correlational features (McDermott et al 2013) can be understood as stages of a cascade of statistical operations (Bruna and Mallat 2013; Andèn and Mallat 2014; Turner and Sahani 2007). The sensory stream may also feed *feature detectors* (e.g., motion), *chunking* mechanisms (Chekaf et al 2016; Christiansen 2018), or specialized processors such as a *face* or *animal detector* (Kirchner and Thorpe 2006; Khayat and Hochstein 2019), or a *speech sound parser* (Christiansen and Chater 2016). Each of these can be understood as a "summary statistic" in a wide sense.

If we agree that new observations are, whenever possible, encoded with reference to existing traces (Cox and Shiffrin 2024), those traces must be found in memory. In particular the detection and discovery of patterns over time implies intensive recourse to a buffer or hierarchy of buffers, that are known to hold detailed sensory information over remarkably long times (Agus et al 2010; Bianco et al 2020; Bastug et al 2026). The input stream being incessant, encoding must operate continuously. The memory-intensive requirements of sensory encoding underscore the importance of efficient search (see below).

## Time

Time and memory are intimately related (Augustine 397; Hooke 1682; Morton et al 2017; Poly and Cutler 2017; Clewett and Davachi 2017; Fountas et al 2022). Temporal order, which reflects causal relations worth remembering, is conserved in memory at least partially (Farrell 2012; Stern et al 2020; Hintzman 2016; D'Argembeau 2020). Sampling of the observation stream might however be non-uniform (D'Argembeau et al 2018), rhythmed by event boundaries (Clewett et al 2019), and in some situations the temporal order might be lost (Fouquet et al 2010).

An influential hypothesis is that of a slowly-varying "context" signal recorded together with observations to detect temporal contiguity (Howard and Kahana 2002; Polyn and Cutler 2017; Eichenbaum 2017). However, there are alternatives that are perhaps simpler. For example, temporal relations can be encoded explicitly with pointers or implicitly by order within an array, as in the time series construct mentioned earlier. Repeated summarization leads to sparser sampling, and thus the remote past is integrated over larger windows, consistent with an overall logarithmically-scaled timeline (Brown et al 2007; Sommer 2016; Howard 2018; Singh et al 2018; Sadeh and Pertzov 2020; Scofield and Johnson 2022). If the context itself requires remembering, its trace can be modeled as a second time series sampled more sparsely than the first, possibly with pointers between the two that account for "context effects" or conditioned responses. Non-uniform sampling might then account for rapid changes of the otherwise-slowly varying context (Dubrow et al 2017).

An alternative to a progressively aggregated time series is a bank of leaky integrators with different time constants. A sensory stream could be coded as a weighted sum of these "basis functions," as in the Laplace transform (Howard 2018; Singh et al 2018; Tiganj et al 2019). A serial representation (as favored in this paper) is more flexible and easier to describe, however a biological implementation might resort to either or both strategies. Indeed, there is evidence for a diversity of time constants, both at the neuronal level (Bernaccia et al 2011; Cavanaugh et al 2020) and within and across cortical areas (Hasson et al 2015; Baldassano et al 2017; Raut et al 2020; Norman-Haignière 2022; Tarder-Stoll 2024), longer time constants being associated with greater abstraction.

The empirical shape of the "forgetting curve" of memories over time has played an important role in memory studies. It is often described as following a *power law*, which might seem inconsistent with the exponential trends referred to earlier. However, a power law can result from averaging over multiple exponential processes (Murre and Chessa 2011), or if storage space is allowed to increase logarithmically with time (Bialek et al 2001), and the empirical power law itself has been questioned (Radavansky et al 2022).

## Chunks and events

If memory traces are discrete, the continuous sensory stream must at some point segmented into chunks or "events" (Zacks et al 2007; Kurby and Zacks 2007; Farell 2012; Chekaf et al 2016; Ben-Yakov and Hensen 2018). According to Event Segmentation Theory (EST, Zacks 2020), events are segments of observations with homogenous statistical properties separated by boundaries determined on the basis of surprise, or particular occurrences such as crossing a doorway (Radavansky and Zacks 2017; Logie and Donaldson 2021). It has been suggested that events organized as a hierarchy over different time scales might serve as an index to facilitate access to subsets of experience (Kurby and Zacks 2008; Baldassano et al 2017).

A chunk might consist of a set of contiguous samples of a time series with a statistical distribution distinct from neighboring chunks (otherwise their representations are redundant). Determining boundaries between chunks requires a data-driven segmentation process that can be either be *on-line* (unidirectional, using observations preceding the boundary, e.g., Aggarwal et al 2007), or *off-line* (bidirectional, using observations on both sides of the boundary). Models of segmentation for memory are mostly of the on-line flavor. Typically, an estimate of *surprise* (also called "surprisal") of a new observation relative to the statistics of the current chunk is used to trigger segmentation (Zacks et al 2007; Baldi and Itti 2010). *Predictive coding* (Palmer et al 2015; Barron et al 2020) may be recruited for the same purpose: observations that poorly fit the current predictive model trigger a new chunk (Kroes and Fernández 2012; Bein et al 2023).

On-line segmentation is appropriate if the data rate is high, but off-line segmentation allows decisions that are better informed (Karp 1992; Nagy and Orban 2016). For example, online K-means clustering can be improved by maintaining a buffer of smaller clusters so that cluster boundaries can be revised as new data come in (Ackerman and Dasgupta 2014). A plausible strategy is to initially place boundaries between all observations and remove them incrementally to form chunks (aggregative clustering). A boundary is prioritized for removal if it separates chunks that follow a similar statistical distribution, which can be detected by applying KL divergence (heuristic #2). Heuristics #3-10 might also contribute to such decisions. Offline segmentation requires a buffer to hold the stream of observations temporarily, which is fortunately what memory itself provides.

Aggregation over time cannot remove redundancies between distant chunks. To efficiently code a recurring pattern requires storing it as an entry in a dictionary and coding its occurrences by *pointers* to that entry. To save space and increase the "hit rate," the entry can consist of a statistical representation that defines a distribution of patterns judged similar. A plausible strategy is to initially assign each observation to a new entry and subsequently cluster entries, guided by the heuristics mentioned earlier. Space permitting, a proportion of entries can be devoted to *outliers*, allowing them to be temporarily discounted from the calculation of statistics (Haberman and Whitney 2010; de Gardelle and Summerfield 2011; Epstein et al 2020), or encoded individually (Brady and Alvarez 2011; Brady et al 2013; Avci and Boduroglu 2021). The idea that modes of a multimodal distribution are encoded separately and later merged (Chetverikov et al 2017b) is consistent with the cluster of samples (CoS) model of Sun et al (2019) and the probabilistic clustering theory (PCT) of Orhan and Jacobs (2013). Such "lazy" clustering addresses the debate between representation by centroids, prototypes (or support vectors), or exemplars (Dubé and Sekuler 2015): all are supported.

## Search

Memory access requires *search*. To trigger a rule that leads to action (as in a behaviorist model), the antecedent of the rule must be found within a store of rules. Search is also required to recognize that an observation is familiar, or recall an event based on a that observation (Richard Semon's "ecphory", Tonegawa et al 2015). To detect repetition requires that each chunk of the input stream be compared with previous observations. To code an event based on a preexisting schema (e.g., Raaijmakers and Shifrin 1981) requires finding that schema among many. To merge pairs of nodes that are similar requires that such pairs be identified, and so on. This implies a massive recourse to search for encoding, retrieval, and curation of memory. With few exceptions (e.g., Kahana 2020; Drugowitsch and Pouget 2024) little attention has been devoted to the issue of *computational complexity* of search, a tacit assumption being that "global search" is somehow possible within a massively parallel brain (Clark and Gronlund 1996).

For visual search, it is common to distinguish *parallel search*, with computational cost close to $O(1)$, from *serial search* with cost $O(N)$ (Tsotsos 1990; Wolfe 2012; Wolfe 2020). Both costs have been observed experimentally for memory (Kahana 2020), but computational arguments tell us that cost $O(1)$ cannot be maintained in the limit of a large memory (Rae 2021; Tsotsos 2022). The best we can hope for is $O(logN)$, which interestingly is observed for hybrid visual search (Wolfe 2012), and for recognition response times (Bright et al 2022). To achieve a cost on this order requires a *hierarchical index* (Tsotsos 1990; Rae 2021), that might be built by applying statistical summarization repeatedly while keeping the intermediate steps. Heuristic #2 would ensure that nodes remain well-separated in terms of KL divergence, as required for such an index. Response times also depend strongly on priming, an effect which might be explained if the index structure were reconfigured after each search to place the matching node at the root (entry point). A computer-science model of this is the *splay tree* (Sleator and Tarjan 1985).

Successful search might return a "found it" flag signaling *familiarity* (Yonelinas 2010). Search using a subset of dimensions as a key might return the full record via "pattern completion" (Marr 1970) leading to *recall*, possibly augmented by its temporal or spatial context ("*context retrieval*", Polyn and Cutler 2017) or semantic context ("*association*", Raajmakers and Shiffrin 1981). Familiarity, recall, encoding, and memory maintenance all entail search. By supporting statistical indexing, the memory-as-statistics hypothesis helps address its needs.

## The memory curation process

The idea of a process been put forward repeatedly (James 1890; Burnham 1903; Bartlett 1932; Lechner et al 1999; McKenzie and Eichenbaum 2011; Dudai 2012; Nadel et al 2012; Hasson et al 2015; Richards and Frankland 2017). It is sometimes described as one-off process of *consolidation*, for example involving transfer from hippocampus to cortex (Lechner et al 1999), or a few-off process of *reconsolidation* after retrieval or updating (McKenzie and Eichenbaum 2011; Nadel et al 2012; Rodriguez-Ortiz et al 2017, Robertson 2018). Dudai (2012) proposed that reconsolidation might occur repeatedly throughout the lifetime, similar to what is proposed here (see also Robertson 2009). Consolidation carries a connotation of enhancement, but the data processing inequality tells us that there can be no increase in information (MacKay 2003). Rather, there is a selective loss that is irreversible: an observation cannot be reinstated from an abstract memory trace, other than by hallucinatory reconstruction (contrast for example with Roüast and Schönauer 2023 who suggest reversibility).

Memory curation involves replacing old traces by more concise new traces, elaborated from the old. Conceptually, it is useful to distinguish three parallel "threads" in this process: a *productive* thread that elaborates a new trace by abstracting the present content together with recent input, a *selective* (or attentive) thread that prioritizes certain traces and tags others for deletion, and a *destructive* thread that carries out the deletion on the basis of the tags. The advantage of distinguishing the last two threads is that deletion can be delayed until the very last moment, at which point it can proceed rapidly to address an urgent need for space. Delayed deletion allows priorities to be revised based on the most recent input which, according to heuristic #1, is likely to be the most useful. While conceptually distinct, these threads might be implemented as a single process at low levels (e.g., sensory) where the rate is fast. At higher levels they might be distinct. The first thread, productive, might include planning, mind-wandering, or creative activities, whereas the second thread, selective, allows less useful creations to be pruned. The distinction between the first two threads maps roughly to that between construction and evaluation in creative ideation (Benedek et al 2023; Ivancovsky et al 2024). That between the last two threads (tagging and deletion) finds some support in tagging mechanisms found at the synaptic level (Redondo and Morris 2011; Dunsmoor et al 2022).

Except for occasional interactions with encoding or retrieval processes, memory curation operates autonomously, converting memory into memory. If we accept that it operates concurrently at multiple scales from concrete to abstract, most of it must be subconscious, with consciously-accessible traces (as in mind wandering) forming the "tip of the iceberg". Curation applies known rules (exploitation), and discovers new ones (exploration). The latter might require the brain's processing powers to "turn inwards" so as to grasp many potential strands of a solution and experiment with their configurations, hence its association with sleep, dreams and mind-wandering. The benefit is compression, but also discovery of potentially useful regularities of a higher order, or "shortcuts in a cognitive map" (Gable et al 2019; Shearer et al 2025). Schmidhuber (2009) suggested an innate drive to "discover new rules to compress data," which might explain *why*, and not just how, Funes found similarities between clouds, foam and book covers.

## Neural basis

It is traditional to associate memory systems (e.g., episodic vs semantic) with different parts of the brain (e.g. Rubin 2006; Kumaran et al 2016; Sekeres et al 2018; Gilboa and Moscovich 2021; see Sherman et al 2024 for a critique). The concrete-to-abstract continuum of MAP does not encourage anatomical parcellation, but the different requirements of fast sensory processing and slower abstract processing might imply anatomical specializations between and within and between regions (e.g., Hasson et al 2015; Irish and Vatansever 2020; Norman-Haignière et al 2022; Sabat et al 2025). Memory has been associated particularly with the hippocampus and specific neocortical structures, several of which belong to the *default mode network* (DMN) of nodes that are less active during tasks than at rest (Buckner et al 2008; Raichle 2010, 2015; Christoff et al 2016; Addis 2018; Alves et al 2019; Menon 2023).

A classic view is that fresh traces reside in hippocampus, and long term "consolidated" traces in cortex (Squire and Alvarez 1995), but recent proposals place the bulk of storage in cortex, with the hippocampus providing an index to the memory trace (Goode et al 2020), or alternatively processing services to write, read or reorganize the trace (Kroes and Fernández 2012; Kumaran et al 2016), over both short and long timescales (Sadeh and Pertzov 2020). MAP does not make any anatomical assumptions, but it is tempting to speculate that the process recruits hippocampus together with other nodes of the DMN, memory curation requiring an ongoing exchange of information between these nodes.

For MAP, an engram parametrizes (according to some statistcal model) a distribution over a space of sensory (or more concrete mnemonic) traces. This raises the question of how such distributions (or their parameters) are represented by neurons (Fiser et al 2010; Pouget et al 2013; Orbán et al 2016; Sanborn 2017; Haefner et al 2024), both for long term storage, and for short-term computation. Storage requires stability and concision, whereas computation requires the ability to expand the trace as a distribution in order to perform inference, compare traces via KL divergence, or perform generative sampling. How these requirements might be satisfied by engram candidates such as synaptic weights or neural activity remains to be worked out, as does the question of translation between storage and computational traces.

## Metabolic cost

The memory process is likely to have a significant metabolic cost. For example, contextual encoding of new observations (e.g., Hasson et al 2015; Zacks 2020) requires that memory be searched, repeatedly and at a high rate. Search based on statistics tied to nodes of a memory structure implies their translation from a storage format to a computational format. If the latter involves neural activity, e.g., a spike-rate-based (Ma et al 2006; Zhang et al 2023) or time-based (Yiling et al 2023; Asabuki and Fukai 2025) population code, search is expected to have a large metabolic cost. Memory maintenance might likewise require the implementation of a statistical measure such KL-divergence (Eq. 1) which might also involve sampling (Mumford 1992; Sanborn and Chater 2016; Orhan et al 2016; Zhang et al 2023), replay, or simulation (Christoff 2016; Addis 2018), all of which are likely expensive.

If a shared pool of computational resources is used to *encode* new observations into traces, *consolidate* them into more abstract traces, *plan* future actions, *act*, and consciously *reflect*, we can expect competition between these various activities (Sherman and Turk-Brown 2020). One might conjecture that sleep, which plays a well-known role in memory consolidation (Diekelman and Born 2010; Vorster and Born 2015), frees resources otherwise tied up in encoding, planning, or action, making them available for certain steps of memory curation. Dreams then might reflect the lingering traces of such curation, for example *generative sampling* of abstract entities to support their reorganization (Llewellyn 2013; Mildner and Tamir 2019), or *replay* of detailed episodes to extract their gist (Domhoff et al 2011; Kumaran et al 2016), or *simulation* of future or counterfactual situations (Addis 2018; Menon 2023).

The brain consumes a disproportionate share of metabolic resources (20% of total for 2% of body weight), and this consumption varies little when a subject performs a task or is at rest (Raichle 2010). Indeed, the DMN, often associated with memory, is *less* active during mental activity or behavior, possibly because the recruitment of certain of its nodes for active thought or behavior leads to reduced activity within the rest of the network (see Wixted 2004 for a similar argument). Whereas summarization of sensory input must be carried out promptly, that of higher-level representations can be spread over time, consistent with a steady metabolic consumption. It is tempting to speculate that the baseline activity of the brain, otherwise hard to explain, is largely devoted to the memory process. What better is there for a brain to do, at rest, than ruminate the information that it contains?

This concludes the review of memory science in the light of the MAP framework. The review was broad but shallow: each topic deserves a much deeper discussion than fits in this paper. The greatest enigma is how the high-rate sensory stream is converted into a persistent trace over a lifetime, overcoming issues of limited storage space, search, and uneven and fading relevance. This enigma, hinted at by Borges but rarely evoked in memory studies, is what MAP aims to address.

# IV. Discussion

Memory underlies every aspect of cognition. To compare two tones in a psychophysical task, we store the trace of the first and compare it with the second, in addition to remembering task instructions (Demany and Semal 2007). Decisions and plans, once made, are held until they are executed, and planification itself benefits from knowledge of the past. Detailed sensory traces stored over extended periods allow us to discover interesting temporal patterns. All of this involves memory.

## Past and future

Memory records the *past* in service of the *future*. These concepts seem straightforward, but we run into problems when thinking of a device in which they are embodied. As an example, the information-theoretic analysis of Bialek et al. (2001) quantifies the amount of mutual information between the past and the future, but to apply that analysis we must conceptually "freeze" time and by the time the results are available time has moved on. This conundrum is captured by another story of Borges, *The Secret Miracle*, in which the protagonist, failed playwright Jaromir Hladik, faces the firing squad but prays for time enough to finish his last piece. His prayer is answered: the bullets freeze in mid-air for one whole year (in his mind), allowing him to rewrite the play to perfection, before the bullets resume their course. Like Jaromir, we might wish to ponder the past and prepare for the future but – except miracle – *we don't have time*: a new past is born at every instant (more relevant than any previous "past"), and a new future. Memory, past, present and future are moving targets.

To escape this conundrum requires distinguishing physical and mental time, as clarified more than 16 centuries ago by Augustine (397, book 11, ch. 20). He noted that "*neither past nor future are in existence*", the first because it *no longer exists* in a physical sense, the second because it *does not yet exist*. And yet, we use these words to good effect so their referents must exist, and indeed they do exist *in the "present" of our minds*. In Augustine's words: "*the present of things past is memory, the present of things present is vision* [i.e., sensory input]*, the present of things future is expectation.*" Past, present and future are mental constructs that our mind creates at some point of time and maintains over some period. They are generated on the basis of information at hand (i.e., memory) the same way that a recollection is generated from abstract memory traces. They speak of physical time, but do not coincide with it. Augustine's distinction between *mental* time and *physical* time neatly sidesteps the paradoxes that occupied later thinkers such as Brentano, Husserl, and others (Husserl 1964; Fréchette 2017). It does so by cleanly separating (and thus allowing their relations to be understood) the physical dynamics of the world, the algorithmic dynamics of the memory trace, and the words that we use to talk about time.

## Past or future?

It has been suggested that DMN activity at rest, during mind wandering or sleep, serves to *plan* or predict the future (Molinaro and Bar 2025), rather than ruminate the past. Is it conceivable that memory is coded as plans, rather than statistics? A plan distills information into a concise, ready-for-use format. Past and future are symmetric in terms of information (Tishby 1999; Bialek et al 2001), and remembering shares much with predicting or planning (Glenberg 1997; Schacter and Addis 2007, Schacter et al 2012, Gershman 2017, Nobre and Stokes 2019, Zacks et al 2020, Josselyn and Tonegawa 2020; Addis and Wagner 2024). Rather than recording the past over multiple time scales, memory could be planning the future at multiple horizons (Bar 2009; Kroes and Fernandez 2012; Bein et al 2023). Rather than ruminating, the brain could be scheming.

If the goal is to store past information, plans are not an ideal storage format, first because we tend to plan for a small number of contingencies and thus the total information content is limited, second because specialization for specific goals is potentially detrimental to other goals (Wolpert 1996), third because there is no clear mechanism by which plans at time $t$ can be merged with new observations to obtain plans at time $t+1$:

$$\mathcal{O}_t, \mathcal{P}_t \rightarrow \mathcal{P}_{t+1} \qquad (3)$$

where $\mathcal{P}_t$ is the set of plans and $\mathcal{O}_t$ are observations (compare with Eq. 2). Plans are elaborated from past observations; if those were available plans would be redundant; that ideal cannot be attained but it is worth approximating with a general-purpose detailed summary of past observations (Lengyel and Dayan 2007; Schmidhuber 2009). That said, a plan or decision might sometimes be an appropriate “statistic” to represent the distribution of observations that triggered it (as when we remember our intention to do something, but not why).

## The sensation-inaction loop

Links between sensation and action have been emphasized (Braitenberg 1984; O'Regan and Noe 2001; Hommel 2005; Cisek 2019; Buzsáki and Tingley 2023), but the most common outcome of stimulation is *inaction*. Most events within our "blooming and buzzing" environment (James 1890) are not worthy of action, or even of attention. Sensory stimulation is dense and behavior sparse (Salge et al 2014), and thus we ignore much more often than we act. The class of stimuli *not* worthy of action or attention has a complex statistical structure that it is useful to acquire so that we can reliably ignore them, as false alarms waste resources. This is arguably a prime goal of memory. Related ideas are *adaptation*, *habituation*, and *predictive coding*.

## Empowerment

"*Two or three times, Funes had reconstructed a whole day,*" but "*each reconstruction had required a whole day.*" This awesome feat implies, a contrario, that most of Ireneo's vast memory was not revisited in detail. Librarians have a similar experience of curating books that no one borrows, and it is tempting to conclude that an unvisited memory, or never-borrowed book, has no value. That would be an error as its value derives from its *potential* usefulness, as captured by the concept of *empowerment* (Klyubin et al 2005; Salge et al 2014). Memory, like money or political clout, expands the range of possible actions and potential rewards, regardless of whether they are carried out or reaped.

## What can we do with an understanding of memory?

Science and technology have long been a source of metaphors for memory (Hooke 1682; Draisma 1995), and neural networks in particular have played a role as models of memory and cognition (e.g., McClelland et al 1995), a role lately taken over by deep neural networks (Graves et al 2016; Kumaran et al 2016; Hassabis et al 2017; Botvinik et al 2019; Richards et al 2019). Conversely, technology has been usefully inspired by knowledge of the brain, and in particular memory. Doing is a touchstone for knowing.

A neural network "remembers" its training data (including supervision signals and reward) distilled into its weights. It also remembers recent input in the pattern of activity across nodes. The "long-term" capacity of the network is determined by the number weights (parameters), and its "short-term" capacity by the number of nodes. The temporal span of its "long-term memory" is function of the training data, modulated by pretraining, retraining, updating, etc., and the span of the short-term part depends on the format of the input buffer and the internal dynamics of the network, in particular in the case of a recurrent neural network (RNN).

Recent progress has been made by augmenting neural networks with additional memory, starting from the influential Long Short-Term Memory (LSTM, Hochreiter and Schmidhuber 1997) and continuing with more recent *memory-augmented* networks (e.g., Gemici et al 2016; Rae et al 2019; Devlin et al 2019; Banino et al 2020; Vaswani et al 2020; Fountas 2024; Behrouz et al 2025), that complement the exponential forgetting of an RNN with an explicit "sample and hold" principle that allows the network to "shop" for information within a wider temporal context. The memory capacity of a classic network (number of parameters) must be limited to avoid overfitting, but an external memory is less constrained. Parameters of a classic network are optimized for the training task, but external memory can be more "general-purpose" (Lengyel and Dayan 2007).

The present framework might offer three contributions. First, it suggests how to design a memory buffer that is *anchored on the present*. Second, it suggests how to increase the *temporal span* of that buffer without increasing its size (the cost of which can be quadratic, e.g., Fountas 2024). Third, it suggests a set of heuristics to reduce the dimensionality of this buffer while retaining value. Search cost is linear in dimension $d$ for approximate nearest neighbor search, exponential for accurate nearest neighbor search (Rae 2021), statistical distributions are expensive to sample in a high-dimensional space, and learning benefits from simple, low-dimensional features (Feldman 2016). Heuristics implement *inductive bias* (Richards et al 2019; Székely et al 2024) analogous to weight-tying in a convolutive neural network, or transfer learning (Delétang et al 2023; Székely et al 2024). There choice of heuristics, and thus of inductive bias, is a potential target for learning or meta-learning (Binz et al 2024).

### In sum

Memory is a store and a process; its role both to retain and to forget. It operates at an urgent pace to record high-bandwidth sensory input and ensure prompt access for timely action, and at a leisurely pace to refine traces conserved over a life time. Traces are abstract and gist-like, yet generative recall mechanisms provide the illusion of concrete observations and events, allowing us to think about them, plan, and communicate our "memories" to others. Memories might seem to defy quantification (as in Proust's memory of biting into a madeleine cake dipped in tea and the reminiscences that it triggered) yet it helps to view them as statistical in nature, requiring tools such as generative sampling, inference, and KL divergence. Much more information is lost than retained. Memory subsumes learning, but goes beyond it by catering to yet unknown needs rather than to a specific task. Memory is all that remains of the past.

The constructive, restless, and forgetful nature of memory is captured by these lines of Andrew Marvel: "*The mind, that ocean where each kind / Does straight its own resemblance find; / Yet it creates, transcending these, / Far other worlds, and other seas; / Annihilating all that's made / To a green thought in a green shade*".

## Conclusion

A rational framework was proposed to explain how we accommodate unbounded sensory input within bounded memory. Memory is stored as statistics organized into structures that are repeatedly summarized and compressed to make room for new input. Repeated summarization requires an intensive ongoing process guided by heuristics that help optimize the memory for future needs. The framework differs from previous accounts of memory by its emphasis on a process that is intensive, complex, and expensive, its reliance on statistics as a representation of memory, and the use of heuristics to guide the choice of statistics at each summarization step. The metabolic requirements of the process might explain the high metabolic consumption of the brain at rest. The framework is intended as an aid to make sense of our extensive knowledge of memory, and bring us closer to an understanding of memory in functional and mechanistic terms.

## Acknowledgments

The work was supported by grants ANR-10-LABX-0087 IEC and ANR-17-EURE-0017 et ANR-10-IDEX-0001-02 awarded to PSL (Université Paris Sciences & Lettres). An early phase received support from the High Council for Scientific and Technological Cooperation between France-Israel (2012-2013). Previous versions of this manuscript benefited from comments by Eli Nelken and Malcolm Slaney, as well as earlier discussions with Tali Tishby. Initial work on scalable statistics was performed at Ircam (Paris), and some material was presented at the 2025 Coghear workshop. No AI tools were used in the preparation.

## References

*[To the reviewers: this list includes items no longer cited in the text. It will be pruned in the final revision.]*

Abbott, J. T., Austerweil, J. L., & Griffiths, T. L. (2012). Human memory search as a random walk in a semantic network. *Advances in Neural Information Processing Systems*. https://proceedings.neurips.cc/paper_files/paper/2012/file/14d9e8007c9b41f57891c48e07c23f57-Paper.pdf

Ackerman, M., & Dasgupta, S. (2014). Incremental Clustering: The Case for Extra Clusters. *Advances in Neural Information Processing Systems*, *27*, 9.

Addis, D. R. (2018). Are episodic memories special? On the sameness of remembered and imagined event simulation. *Journal of the Royal Society of New Zealand*, *48*(2–3), 64–88. https://doi.org/10.1080/03036758.2018.1439071

Aggarwal CC (2007) Data Streams: Models and Algorithms Models and Algorithms, Springer Science & Business Media.

Agus, T. R., Thorpe, S. J., & Pressnitzer, D. (2010). Rapid Formation of Robust Auditory Memories: Insights from Noise. *Neuron*, *66*(4), 610–618. https://doi.org/10.1016/j.neuron.2010.04.014

Alves, P. N., Foulon, C., Karolis, V., Bzdok, D., Margulies, D. S., Volle, E., & Thiebaut De Schotten, M. (2019). An improved neuroanatomical model of the default-mode network reconciles previous neuroimaging and neuropathological findings. *Communications Biology*, *2*(1), 370. https://doi.org/10.1038/s42003-019-0611-3

Andén J, Mallat S (2014) Deep scattering spectrum. IEEE Transactions on Signal Processing 62:4114 – 4128.

Anderson, J. R. (1983). A Spreading Activation Theory of Memory. *Journal of Verbal Learning and Verbal Behavior*, *22*, 261–295.

Anderson, J. R., & Milson, R. (1989). Human Memory: An Adaptive Perspective. *Psychological Review*, *96*, 703–719.

Anderson, J. R., & Schooler, L. J. (1991). Reflections of the Environment in Memory. *Psychological Science*, *2*(6), 396–408. https://doi.org/10.1111/j.1467-9280.1991.tb00174.x

Ariely D (2001) Seeing Sets: Representation by Statistical Properties. Psychological Science 12:157–162.

Attneave, F. (1954). Some informational aspects of visual perception. *Psychological Review*, *61*(3), 183–193. https://doi.org/10.1037/h0054663

Augustine (circa 397) Confessions, book 10, chapters 9-21 on memory, reviewed by Cassel (2012), book 11, chapters 17-28 on time, reviewed by Manning (2013).

Avci, B., & Boduroglu, A. (2021). Contributions of ensemble perception to outlier representation precision. *Attention, Perception, & Psychophysics*, *83*(3), 1141–1151. https://doi.org/10.3758/s13414-021-02270-9

Awh, E., Belopolsky, A. V., & Theeuwes, J. (2012). Top-down versus bottom-up attentional control: A failed theoretical dichotomy. *Trends in Cognitive Sciences*, *16*(8), 437–443. https://doi.org/10.1016/j.tics.2012.06.010

Baddeley, A. D., & Hitch, G. J. (2019). The phonological loop as a buffer store: An update. *Cortex*, *112*, 91–106. https://doi.org/10.1016/j.cortex.2018.05.015

Bainbridge, W. A. (2020). The resiliency of image memorability: A predictor of memory separate from attention and priming. *Neuropsychologia*, *141*, 107408. https://doi.org/10.1016/j.neuropsychologia.2020.107408

Baldassano, C., Chen, J., Zadbood, A., Pillow, J. W., Hasson, U., & Norman, K. A. (2017). Discovering Event Structure in Continuous Narrative Perception and Memory. *Neuron*, *95*(3), 709-721.e5. https://doi.org/10.1016/j.neuron.2017.06.041

Baldi, Pierre, and Laurent Itti. 2010. 'Of Bits and Wows: A Bayesian Theory of Surprise with Applications to Attention'. *Neural Networks* 23 (5): 649–66. https://doi.org/10.1016/j.neunet.2009.12.007.

Banino, A., Badia, A. P., Köster, R., Chadwick, M. J., Zambaldi, V., Hassabis, D., Barry, C., Botvinick, M., Kumaran, D., & Blundell, C. (2020). *MEMO: A Deep Network for Flexible Combination of Episodic Memories* (arXiv:2001.10913). arXiv. https://doi.org/10.48550/arXiv.2001.10913

Barlow, H. B. (1959). Sensory mechanisms, the reduction of redundancy and intelligence. In D. V. Blake & A. M. Uttley (Eds.), Proceedings of the Symposium on the Mechanization of Thought Processes, (Vol. 2, pp. 537–574), London: H. M. Stationery Office.

Barlow, Horace. 1990. 'Conditions for Versatile Learning, Helmholtz's Unconscious Inference, and the Task of Perception'. *Vision Research* 30 (11): 1561–71. https://doi.org/10.1016/0042-6989(90)90144-A.

Barron, H. C., Auksztulewicz, R., & Friston, K. (2020). Prediction and memory: A predictive coding account. *Progress in Neurobiology*, *192*, 101821. https://doi.org/10.1016/j.pneurobio.2020.101821

Barry, D. N., & Maguire, E. A. (2019). Remote Memory and the Hippocampus: A Constructive Critique. *Trends in Cognitive Sciences*, *23*(2), 128–142. https://doi.org/10.1016/j.tics.2018.11.005

Barsalou, L. W. (1999). Perceptual symbol systems. *Behavioral and Brain Sciences*, *22*(4), 577–660. https://doi.org/10.1017/S0140525X99002149

Bartlett, F.C., 1932. Remembering: A Study in Experimental and Social Psychology. Cambridge University Press.

Bastug, B., Rajendran, V., Bianco, R., Agus, T., Chait, M., & Pressnitzer, D. (2026). Memory for repeated auditory textures. *Cognition*, *268*, 106350. https://doi.org/10.1016/j.cognition.2025.106350

Bates, C. J., & Jacobs, R. A. (2020). Efficient data compression in perception and perceptual memory. *Psychological Review*, *127*(5), 891–917. https://doi.org/10.1037/rev0000197

Behrouz, A., Razaviyayn, M., Zhong, P., & Mirrokni, V. (2025). *It's All Connected: A Journey Through Test-Time Memorization, Attentional Bias, Retention, and Online Optimization* (arXiv:2504.13173). arXiv. https://doi.org/10.48550/arXiv.2504.13173

Bein, O., Gasser, C., Amer, T., Maril, A., & Davachi, L. (2023). Predictions transform memories: How expected versus unexpected events are integrated or separated in memory. *Neuroscience & Biobehavioral Reviews*, *153*, 105368. https://doi.org/10.1016/j.neubiorev.2023.105368

Benedek, M., Beaty, R. E., Schacter, D. L., & Kenett, Y. N. (2023). The role of memory in creative ideation. *Nature Reviews Psychology*, *2*(4), 246–257. https://doi.org/10.1038/s44159-023-00158-z

Ben-Yakov, A., & Henson, R. N. (2018). The Hippocampal Film Editor: Sensitivity and Specificity to Event Boundaries in Continuous Experience. *The Journal of Neuroscience*, *38*(47), 10057–10068. https://doi.org/10.1523/JNEUROSCI.0524-18.2018

Bernacchia, A., Seo, H., Lee, D., & Wang, X.-J. (2011). A reservoir of time constants for memory traces in cortical neurons. *Nature Neuroscience*, *14*(3), 366–372. https://doi.org/10.1038/nn.2752

Bialek, William, Ilya Nemenman, and Naftali Tishby (2001). 'Predictability, Complexity, and Learning'. *Neural Computation* 13 (11): 2409–63. https://doi.org/10.1162/089976601753195969.

Bianco, R., Harrison, P. M., Hu, M., Bolger, C., Picken, S., Pearce, M. T., & Chait, M. (2020). Long-term implicit memory for sequential auditory patterns in humans. *eLife*, *9*, e56073. https://doi.org/10.7554/eLife.56073

Binz, M., Dasgupta, I., Jagadish, A. K., Botvinick, M., Wang, J. X., & Schulz, E. (2024). Meta-learned models of cognition. *Behavioral and Brain Sciences*, *47*, e147. https://doi.org/10.1017/S0140525X23003266

Bloom, B. H. (1970). Space/time trade-offs in hash coding with allowable errors. *Communications of the ACM*, *13*(7), 422–426. https://doi.org/10.1145/362686.362692

Boring EG (1942) Sensation and perception in the history of experimental psychology. New York: Appleton-Century.

Botvinick, M., Ritter, S., Wang, J. X., Kurth-Nelson, Z., Blundell, C., & Hassabis, D. (2019). Reinforcement Learning, Fast and Slow. *Trends in Cognitive Sciences*, *23*(5), 408–422. https://doi.org/10.1016/j.tics.2019.02.006

Brady, T. F., & Alvarez, G. A. (2011). Hierarchical Encoding in Visual Working Memory: Ensemble Statistics Bias Memory for Individual Items. *Psychological Science*, *22*(3), 384–392. https://doi.org/10.1177/0956797610397956

Brady, T. F., Konkle, T., & Alvarez, G. A. (2011). A review of visual memory capacity: Beyond individual items and toward structured representations. *Journal of Vision*, *11*(5), 4–4. https://doi.org/10.1167/11.5.4
Brady, T. F., & Tenenbaum, J. B. (2013). A probabilistic model of visual working memory: Incorporating higher order regularities into working memory capacity estimates. *Psychological Review*, *120*(1), 85–109. https://doi.org/10.1037/a0030779
Brady, T. F., Konkle, T., Gill, J., Oliva, A., & Alvarez, G. A. (2013). Visual Long-Term Memory Has the Same Limit on Fidelity as Visual Working Memory. *Psychological Science*, *24*(6), 981–990. https://doi.org/10.1177/0956797612465439
Brady, T. F., Robinson, M. M., & Williams, J. R. (2024). Noisy and hierarchical visual memory across timescales. *Nature Reviews Psychology*, *3*(3), 147–163. https://doi.org/10.1038/s44159-024-00276-2
Braitenberg, V. (1984). Vehicles: Experiments in Synthetic Psychology. Cambridge, MA: MIT Press, Bradford Books
Braun, E. K., Wimmer, G. E., & Shohamy, D. (2018). Retroactive and graded prioritization of memory by reward. *Nature Communications*, *9*(1). https://doi.org/10.1038/s41467-018-07280-0
Bright, I. M., Singh, I., DiDomenica, R., Oliva, A., & Howard, M. W. (2022). *The time to initiate retrieval of a memory depends on recency*. https://doi.org/10.1101/2022.09.16.508287
Brin, S., & Page, L. (1998). The anatomy of a large-scale hypertextual Web search engine. *Computer Networks and ISDN Systems*, *30*(1–7), 107–117. https://doi.org/10.1016/S0169-7552(98)00110-X
Brown, G. D. A., Neath, I., & Chater, N. (2007). A temporal ratio model of memory. *Psychological Review*, *114*(3), 539–576. https://doi.org/10.1037/0033-295X.114.3.539
Bruna, J., & Mallat, S. (2013). Invariant Scattering Convolution Networks. *IEEE Transactions on Pattern Analysis and Machine Intelligence*, *35*(8), 1872–1886. https://doi.org/10.1109/TPAMI.2012.230
Buckner, R. L., Andrews-Hanna, J. R., & Schacter, D. L. (2008). *The Brain's Default Network: Anatomy, Function, and Relevance to Disease*. *Annals of the New York Academy of Sciences*, *1124*(1), 1–38. https://doi.org/10.1196/annals.1440.011
Burnham, W. H. (1888). Memory, Historically and Experimentally Considered. I. An Historical Sketch of the Older Conceptions of Memory. *The American Journal of Psychology*, *2*(1), 39. https://doi.org/10.2307/1411406
Burnham, Wm. H. (1903). Retroactive Amnesia: Illustrative Cases and a Tentative Explanation. *The American Journal of Psychology*, *14*(3/4), 118. https://doi.org/10.2307/1412310
Burrell, Q. L. (1985). A NOTE ON AGEING IN A LIBRARY CIRCULATION MODEL. *Journal of Documentation*, *41*(2), 100–115. https://doi.org/10.1108/eb026775
Butz, M. V., Achimova, A., Bilkey, D., & Knott, A. (2021). Event-Predictive Cognition: A Root for Conceptual Human Thought. *Topics in Cognitive Science*, *13*(1), 10–24. https://doi.org/10.1111/tops.12522
Buzsáki, G., & Tingley, D. (2023). Cognition from the Body-Brain Partnership: Exaptation of Memory. *Annual Review of Neuroscience*, *46*(1), 191–210. https://doi.org/10.1146/annurev-neuro-101222-110632
Candia, C., Jara-Figueroa, C., Rodriguez-Sickert, C., Barabási, A.-L., & Hidalgo, C. A. (2019). The universal decay of collective memory and attention. *Nature Human Behaviour*, *3*(1), 82–91. https://doi.org/10.1038/s41562-018-0474-5
Cassel, J.-C., Cassel, D., & Manning, L. (2012). From Augustine of Hippo's Memory Systems to Our Modern Taxonomy in Cognitive Psychology and Neuroscience of Memory: A 16-Century Nap of Intuition before Light of Evidence. *Behavioral Sciences*, *3*(1), 21–41. https://doi.org/10.3390/bs3010021
Cavanagh, S. E., Hunt, L. T., & Kennerley, S. W. (2020). A Diversity of Intrinsic Timescales Underlie Neural Computations. *Frontiers in Neural Circuits*, *14*, 615626. https://doi.org/10.3389/fncir.2020.615626
Chater, Nick & Oaksford, Mike. (1999). Ten years of the rational analysis of cognition. *Trends in Cognitive Sciences*, *3*, 57–65.

Chekaf, M., Cowan, N., & Mathy, F. (2016). Chunk formation in immediate memory and how it relates to data compression. *Cognition*, *155*, 96–107. https://doi.org/10.1016/j.cognition.2016.05.024
Chetverikov, A., Campana, G., & Kristjánsson, Á. (2017a). Learning features in a complex and changing environment: A distribution-based framework for visual attention and vision in general. In *Progress in Brain Research* (Vol. 236, pp. 97–120). Elsevier. https://doi.org/10.1016/bs.pbr.2017.07.001
Chetverikov, A., Campana, G., & Kristjánsson, Á. (2017b). Rapid learning of visual ensembles. *Journal of Vision*, *17*(2), 21. https://doi.org/10.1167/17.2.21
Chklovskii, D. B., Mel, B. W., & Svoboda, K. (2004). Cortical rewiring and information storage. *Nature*, *431*(7010), 782–788. https://doi.org/10.1038/nature03012
Christiansen, M. H., & Chater, N. (2016). The Now-or-Never bottleneck: A fundamental constraint on language. *Behavioral and Brain Sciences*, *39*, 1–72. https://doi.org/10.1017/S0140525X1500031X
Christoff, K., Irving, Z. C., Fox, K. C. R., Spreng, R. N., & Andrews-Hanna, J. R. (2016). Mind-wandering as spontaneous thought: A dynamic framework. *Nature Reviews Neuroscience*, *17*(11), 718–731. https://doi.org/10.1038/nrn.2016.113
Christophel, T. B., Klink, P. C., Spitzer, B., Roelfsema, P. R., & Haynes, J.-D. (2017). The Distributed Nature of Working Memory. *Trends in Cognitive Sciences*, *21*(2), 111–124. https://doi.org/10.1016/j.tics.2016.12.007
Chun, M. M., Golomb, J. D., & Turk-Browne, N. B. (2011). A Taxonomy of External and Internal Attention. *Annual Review of Psychology*, *62*(1), 73–101. https://doi.org/10.1146/annurev.psych.093008.100427
Cisek, P. (2019). Resynthesizing behavior through phylogenetic refinement. *Attention, Perception, & Psychophysics*, *81*(7), 2265–2287. https://doi.org/10.3758/s13414-019-01760-1
Clark, S. E., & Gronlund, S. D. (1996). Global matching models of recognition memory: How the models match the data. *Psychonomic Bulletin & Review*, *3*(1), 37–60. https://doi.org/10.3758/BF03210740
Clark, A. (2013). Whatever next? Predictive brains, situated agents, and the future of cognitive science. *Behavioral and Brain Sciences*, *36*(3), 181–204. https://doi.org/10.1017/S0140525X12000477
Clewett, D., & Davachi, L. (2017). The ebb and flow of experience determines the temporal structure of memory. *Current Opinion in Behavioral Sciences*, *17*, 186–193. https://doi.org/10.1016/j.cobeha.2017.08.013
Clewett, D., DuBrow, S., & Davachi, L. (2019). Transcending time in the brain: How event memories are constructed from experience. *Hippocampus*, *29*(3), 162–183. https://doi.org/10.1002/hipo.23074
Coltheart, M. (1980). Iconic memory and visible persistence. *Perception & Psychophysics*, *27*(3), 183–228. https://doi.org/10.3758/BF03204258
Connerton, P. (2008). Seven types of forgetting. *Memory Studies*, *1*(1), 59–71. https://doi.org/10.1177/1750698007083889
Conway, M. A. (2009). Episodic memories. *Neuropsychologia*, *47*(11), 2305–2313. https://doi.org/10.1016/j.neuropsychologia.2009.02.003
Corballis, M. C. (2019). Mental time travel, language, and evolution. *Neuropsychologia*, *134*, 107202. https://doi.org/10.1016/j.neuropsychologia.2019.107202
Cowan, N. (1984). On Short and Long-Term Auditory Stores. *Psychological Bulletin*, *96*, 341–370.
Cowan, N. (2017). The many faces of working memory and short-term storage. *Psychonomic Bulletin & Review*, *24*(4), 1158–1170. https://doi.org/10.3758/s13423-016-1191-6
Cowan, N. (2019). Short-term memory based on activated long-term memory: A review in response to Norris (2017). *Psychological Bulletin*, *145*(8), 822–847. https://doi.org/10.1037/bul0000199
Cowan, E. T., Chanales, A. J., Davachi, L., & Clewett, D. (2024). Goal Shifts Structure Memories and Prioritize Event-defining Information in Memory. *Journal of Cognitive Neuroscience*, *36*(11), 2415–2431. https://doi.org/10.1162/jocn_a_02220
Cox, G. E. & Shiffrin, R.M. (2024). Computational Models of Event Memory. In Kahana MJ & Wagner AD, *Oxford Handbook of Human Memory*. https://doi.org/10.1093/oxfordhb/9780190917982.013.3

Craik, F. I. M., & Lockhart, R. S. (1972). Levels of processing: A framework for memory research. *Journal of Verbal Learning and Verbal Behavior*, *11*(6), 671–684. https://doi.org/10.1016/S0022-5371(72)80001-X
Craik, FIM (2021) Remembering: an activity of mind and brain, Oxford Academic https://doi.org/10.1093/oso/9780192895226.001.0001.
Cunningham, C. A., Yassa, M. A., & Egeth, H. E. (2015). Massive memory revisited: Limitations on storage capacity for object details in visual long-term memory. *Learning & Memory*, *22*(11), 563–566. https://doi.org/10.1101/lm.039404.115
D'Argembeau, A. (2020). Zooming In and Out on One's Life: Autobiographical Representations at Multiple Time Scales. *Journal of Cognitive Neuroscience*, *32*(11), 2037–2055. https://doi.org/10.1162/jocn_a_01556
D'Argembeau, A., Jeunehomme, O., & Stawarczyk, D. (2022). Slices of the past: How events are temporally compressed in episodic memory. *Memory*, *30*(1), 43–48. https://doi.org/10.1080/09658211.2021.1896737
Darwin, Charles (1857). Letter no. 2130.  Darwin Correspondence Project. https://www.darwinproject.ac.uk/letter?docId=letters/DCP-LETT-2130.xml
Dasgupta, I., Schulz, E., Goodman, N. D., & Gershman, S. J. (2018a). Remembrance of inferences past: Amortization in human hypothesis generation. *Cognition*, *178*, 67–81. https://doi.org/10.1016/j.cognition.2018.04.017
Dasgupta, S., Sheehan, T. C., Stevens, C. F., & Navlakha, S. (2018b). A neural data structure for novelty detection. *Proceedings of the National Academy of Sciences*, *115*(51), 13093–13098. https://doi.org/10.1073/pnas.1814448115
Dau, T., Kollmeier, B., & Kohlrausch, A. (1997). Modeling auditory processing of amplitude modulation. I. Detection and masking with narrow-band carriersa). *J. Acoust. Soc. Am.*, *102*, 2892–2905.
De Brigard, F. (2014a). Is memory for remembering? Recollection as a form of episodic hypothetical thinking. *Synthese*, *191*(2), 155–185. https://doi.org/10.1007/s11229-013-0247-7
De Brigard, F. (2014b). The Nature of Memory Traces. *Philosophy Compass*, *9*(6), 402–414. https://doi.org/10.1111/phc3.12133
De Brigard, F. (2024). Episodic memory without autonoetic consciousness. *Philosophical Transactions of the Royal Society B: Biological Sciences*, *379*(1913), 20230410. https://doi.org/10.1098/rstb.2023.0410
de Cheveigné, A. (2022) Graceful Forgetting II: Data as a Process. arXiv, https://arxiv.org/abs/2211.15441.
de Gardelle, V., & Summerfield, C. (2011). Robust averaging during perceptual judgment. *Proceedings of the National Academy of Sciences*, *108*(32), 13341–13346. https://doi.org/10.1073/pnas.1104517108
Delétang, G., Ruoss, A., Duquenne, P.-A., Catt, E., Genewein, T., Mattern, C., Grau-Moya, J., Wenliang, L. K., Aitchison, M., Orseau, L., Hutter, M., & Veness, J. (2024). *Language Modeling Is Compression* (arXiv:2309.10668). arXiv. https://doi.org/10.48550/arXiv.2309.10668
Demany, L., & Semal, C. (2007). The Role of Memory in Auditory Perception. In W. A. Yost, A. N. Popper, & R. R. Fay (Eds.), *Auditory Perception of Sound Sources* (Vol. 29, pp. 77–113). Springer US. https://doi.org/10.1007/978-0-387-71305-2_4
D'Esposito, M. and Postle, B. R. (2015). The cognitive neuroscience of working memory. Annual Review of Psychology, 66, 115–142.
De Sousa, A. F., Chowdhury, A., & Silva, A. J. (2021). Dimensions and mechanisms of memory organization. *Neuron*, *109*(17), 2649–2662. https://doi.org/10.1016/j.neuron.2021.06.014
Devlin, J., Chang, M.-W., Lee, K., & Toutanova, K. (2019). *BERT: Pre-training of Deep Bidirectional Transformers for Language Understanding* (arXiv:1810.04805). arXiv. https://doi.org/10.48550/arXiv.1810.04805
Diekelmann, S., & Born, J. (2010). The memory function of sleep. *Nature Reviews Neuroscience*, *11*(2), 114–126. https://doi.org/10.1038/nrn2762
Domhoff, G. W. (2011). The neural substrate for dreaming: Is it a subsystem of the default network? *Consciousness and Cognition*, *20*(4), 1163–1174. https://doi.org/10.1016/j.concog.2011.03.001
Draaisma, D. (1995, english translation 2000). Metaphors of memory. Cambridge University Press.

Drugowitsch, J., & Pouget, A. (2024). *Major sources of computational complexity in complex decision-making*. https://doi.org/10.31219/osf.io/vw4yr
Dubé, C., & Sekuler, R. (2015). Obligatory and adaptive averaging in visual short-term memory. *Journal of Vision*, *15*(4), 13. https://doi.org/10.1167/15.4.13
Dudai, Y. (1997). How Big Is Human Memory, or On Being Just Useful Enough. *Learning & Memory*, *3*, 341–365.
Dudai, Y. (2004). The Neurobiology of Consolidations, Or, How Stable is the Engram? *Annual Review of Psychology*, *55*(1), 51–86. https://doi.org/10.1146/annurev.psych.55.090902.142050
Dudai, Y. (2012). The Restless Engram: Consolidations Never End. *Annual Review of Neuroscience*, *35*(1), 227–247. https://doi.org/10.1146/annurev-neuro-062111-150500
Dudai, Y., Karni, A., & Born, J. (2015). The Consolidation and Transformation of Memory. *Neuron*, *88*(1), 20–32. https://doi.org/10.1016/j.neuron.2015.09.004
Dunsmoor, J. E., Murty, V. P., Davachi, L., & Phelps, E. A. (2015). Emotional learning selectively and retroactively strengthens memories for related events. *Nature*, *520*(7547), 345–348. https://doi.org/10.1038/nature14106
Dunsmoor, J. E., Murty, V. P., Clewett, D., Phelps, E. A., & Davachi, L. (2022). Tag and capture: How salient experiences target and rescue nearby events in memory. *Trends in Cognitive Sciences*, *26*(9), 782–795. https://doi.org/10.1016/j.tics.2022.06.009
Duszkiewicz, A. J., McNamara, C. G., Takeuchi, T., & Genzel, L. (2019). Novelty and Dopaminergic Modulation of Memory Persistence: A Tale of Two Systems. *Trends in Neurosciences*, *42*(2), 102–114. https://doi.org/10.1016/j.tins.2018.10.002
Ebbinghaus, H. (1885/1913). Memory: A contribution to experimental psychology. NewYork Teachers College, Columbia University. http://psychclassics.yorku.ca/Ebbinghaus/.
Eco, U, (1988). An Ars Oblivionalis? Forget It! (translated by Marilyn Migiel), *PMLA/Publications of the Modern Language Association of America. 1988;103(3):254-261. doi:10.2307/462374*.
Eichenbaum, H. (2017). Memory: Organization and Control. *Annual Review of Psychology*, *68*(1), 19–45. https://doi.org/10.1146/annurev-psych-010416-044131
Endersby, J. (2009). Lumpers and Splitters: Darwin, Hooker, and the Search for Order. *Science*, *326*(5959), 1496–1499. https://doi.org/10.1126/science.1165915
Endress, A. D., & Potter, M. C. (2014). Large capacity temporary visual memory. *Journal of Experimental Psychology: General*, *143*(2), 548–565. https://doi.org/10.1037/a0033934
Endress, A. D. (2023). In defense of epicycles: Embracing complexity in psychological explanations. *Mind & Language*, *38*(5), 1208–1237. https://doi.org/10.1111/mila.12450
Epstein, M. L., Quilty-Dunn, J., Mandelbaum, E., & Emmanouil, T. A. (2020). The Outlier Paradox: The Role of Iterative Ensemble Coding in Discounting Outliers. *Journal of Experimental Psychology: Human Perception and Performance*, *46*, 1267–1279.
Estes, W. K. (1997). Processes of Memory Loss, Recovery, and Distortion. *Psychological Review*, *104*, 148–169.
Fan, J. E., Hutchinson, J. B., & Turk-Browne, N. B. (2016). When past is present: Substitutions of long-term memory for sensory evidence in perceptual judgments. *Journal of Vision*, *16*(8), 1. https://doi.org/10.1167/16.8.1
Farrell, S. (2012). Temporal clustering and sequencing in short-term memory and episodic memory. *Psychological Review*, *119*(2), 223–271. https://doi.org/10.1037/a0027371
Fawcett, J. M., & Hulbert, J. C. (2020). The Many Faces of Forgetting: Toward a Constructive View of Forgetting in Everyday Life. *Journal of Applied Research in Memory and Cognition*, *9*(1), 1–18. https://doi.org/10.1016/j.jarmac.2019.11.002
Feldman, J. (2016). The simplicity principle in perception and cognition. *WIREs Cognitive Science*, *7*(5), 330–340. https://doi.org/10.1002/wcs.1406
Fernández, J. (2015). What are the benefits of memory distortion? *Consciousness and Cognition*, *33*, 536–547. https://doi.org/10.1016/j.concog.2014.09.019
Fougnie, D., Cormiea, S. M., Zhang, J., Alvarez, G. A., & Wolfe, J. M. (2015). Winter is coming: How humans forage in a temporally structured environment. *Journal of Vision*, *15*(11), 1. https://doi.org/10.1167/15.11.1

Fouquet, C., Tobin, C., & Rondi-Reig, L. (2010). A new approach for modeling episodic memory from rodents to humans: The temporal order memory. *Behavioural Brain Research*, *215*(2), 172–179. https://doi.org/10.1016/j.bbr.2010.05.054
Fountas, Z., Benfeghoul, M. A., Oomerjee, A., Christopoulou, F., Lampouras, G., Bou-Ammar, H., & Wang, J. (2024). *Human-like Episodic Memory for Infinite Context LLMs* (arXiv:2407.09450). arXiv. https://doi.org/10.48550/arXiv.2407.09450
Frankland, P. W., Josselyn, S. A., & Köhler, S. (2019). The neurobiological foundation of memory retrieval. *Nature Neuroscience*, *22*(10), 1576–1585. https://doi.org/10.1038/s41593-019-0493-1
Fréchette, G. (2017). Brentano on Time-Consciousness. In U. Kriegel (Ed.), *The Routledge Handbook of Franz Brentano and the Brentano School* (1st ed., pp. 75–86). Routledge. https://doi.org/10.4324/9781315776460-8
Froyen, V., Feldman, J., & Singh, M. (2015). Bayesian hierarchical grouping: Perceptual grouping as mixture estimation. *Psychological Review*, *122*(4), 575–597. https://doi.org/10.1037/a0039540
Fountas, Z., Sylaidi, A., Nikiforou, K., Seth, A. K., Shanahan, M., & Roseboom, W. (2022). A Predictive Processing Model of Episodic Memory and Time Perception. *Neural Computation*, *34*(7), 1501–1544. https://doi.org/10.1162/neco_a_01514
Gable, S. L., Hopper, E. A., & Schooler, J. W. (2019). When the Muses Strike: Creative Ideas of Physicists and Writers Routinely Occur During Mind Wandering. *Psychological Science*, *30*(3), 396–404. https://doi.org/10.1177/0956797618820626
Gallo, D. A. (2010). False memories and fantastic beliefs: 15 years of the DRM illusion. *Memory & Cognition*, *38*(7), 833–848. https://doi.org/10.3758/MC.38.7.833
Gardner-Medwin, A. R., & Barlow, H. B. (2001). The Limits of Counting Accuracy in Distributed Neural Representations. *Neural Computation*, *13*, 477–504.
Gemici, M., Hung, C.-C., Santoro, A., Wayne, G., Mohamed, S., Rezende, D. J., Amos, D., & Lillicrap, T. (2017). *Generative Temporal Models with Memory* (arXiv:1702.04649). arXiv. https://doi.org/10.48550/arXiv.1702.04649
Gershman, S. J., & Niv, Y. (2010). Learning latent structure: Carving nature at its joints. *Current Opinion in Neurobiology*, *20*(2), 251–256. https://doi.org/10.1016/j.conb.2010.02.008
Gershman, S. J. (2017). Predicting the past, remembering the future. *Current Opinion in Behavioral Sciences*, *17*, 7–13. https://doi.org/10.1016/j.cobeha.2017.05.025
Gershman, S. J., Monfils, M.-H., Norman, K. A., & Niv, Y. (2017). *The computational nature of memory modification*. 41. https://doi.org/10.7554/eLife.23763.001
Gershman, S. J., Radulescu, A., Norman, K. A., & Niv, Y. (2014). Statistical Computations Underlying the Dynamics of Memory Updating. *PLoS Computational Biology*, *10*(11), e1003939. https://doi.org/10.1371/journal.pcbi.1003939
Gershman, S. J. (2024). The Rational Analysis of Memory. In M. J. Kahana & A. D. Wagner (Eds.), *The Oxford Handbook of Human Memory, Two Volume Pack* (1st ed., pp. 1505–1520). Oxford University Press. https://doi.org/10.1093/oxfordhb/9780190917982.013.51
Ghosh, V. E., & Gilboa, A. (2014). What is a memory schema? A historical perspective on current neuroscience literature. *Neuropsychologia*, *53*, 104–114. https://doi.org/10.1016/j.neuropsychologia.2013.11.010
Gilboa, A., & Moscovitch, M. (2021). No consolidation without representation: Correspondence between neural and psychological representations in recent and remote memory. *Neuron*, *109*(14), 2239–2255. https://doi.org/10.1016/j.neuron.2021.04.025
Glenberg, A. M. (1997). What memory is for. *Behavioral and Brain Sciences*, *20*(1), 1–19. https://doi.org/10.1017/S0140525X97000010
Golosovsky, M., & Solomon, S. (2017). Growing complex network of citations of scientific papers—Measurements and modeling. *Physical Review E*, *95*(1), 012324. https://doi.org/10.1103/PhysRevE.95.012324
Goode, T. D., Tanaka, K. Z., Sahay, A., & McHugh, T. J. (2020). An Integrated Index: Engrams, Place Cells, and Hippocampal Memory. *Neuron*, *107*(5), 805–820. https://doi.org/10.1016/j.neuron.2020.07.011
Graves, A., Wayne, G., Reynolds, M., Harley, T., Danihelka, I., Grabska-Barwińska, A., Colmenarejo, S. G., Grefenstette, E., Ramalho, T., Agapiou, J., Badia, A. P., Hermann, K. M., Zwols, Y., Ostrovski, G., Cain, A., King, H., Summerfield, C., Blunsom, P., Kavukcuoglu, K., &

Hassabis, D. (2016). Hybrid computing using a neural network with dynamic external memory. *Nature*, *538*(7626), 471–476. https://doi.org/10.1038/nature20101
Graziano, M., & Sigman, M. (2008). The dynamics of sensory buffers: Geometric, spatial, and experience-dependent shaping of iconic memory. *Journal of Vision*, *8*(5), 9. https://doi.org/10.1167/8.5.9
Griffiths, T. L., Steyvers, M., & Firl, A. (2007). Google and the Mind: Predicting Fluency With PageRank. *Psychological Science*, *18*(12), 1069–1076. https://doi.org/10.1111/j.1467-9280.2007.02027.x
Griffiths, T. L., Steyvers, M., & Tenenbaum, J. B. (2007). Topics in semantic representation. *Psychological Review*, *114*(2), 211–244. https://doi.org/10.1037/0033-295X.114.2.211
Ha, D., & Schmidhuber, J. (2018). *World Models*. https://doi.org/10.5281/zenodo.1207631
Haberman, J., & Whitney, D. (2007). Rapid extraction of mean emotion and gender from sets of faces. *Current Biology*, *17*(17), R751–R753. https://doi.org/10.1016/j.cub.2007.06.039
Haberman, J., Harp, T., & Whitney, D. (2009). Averaging facial expression over time. *Journal of Vision*, *9*(11), 1–13. https://doi.org/10.1167/9.11.1
Haberman, J., & Whitney, D. (2009). Seeing the mean: Ensemble coding for sets of faces. *Journal of Experimental Psychology: Human Perception and Performance*, *35*(3), 718–734. https://doi.org/10.1037/a0013899
Haberman, J., & Whitney, D. (2010). The visual system discounts emotional deviants when extracting average expression. *Attention, Perception & Psychophysics*, *72*(7), 1825–1838. https://doi.org/10.3758/APP.72.7.1825
Haefner, R. M., Beck, J., Savin, C., Salmasi, M., & Pitkow, X. (2024). *How does the brain compute with probabilities?* (arXiv:2409.02709). arXiv. https://doi.org/10.48550/arXiv.2409.02709
Hardt, O., Nader, K., & Nadel, L. (2013). Decay happens: The role of active forgetting in memory. *Trends in Cognitive Sciences*, *17*(3), 111–120. https://doi.org/10.1016/j.tics.2013.01.001
Hassabis, D., & Maguire, E. A. (2009). The construction system of the brain. *Philosophical Transactions of the Royal Society B: Biological Sciences*, *364*(1521), 1263–1271. https://doi.org/10.1098/rstb.2008.0296
Hassabis, D., Kumaran, D., Summerfield, C., & Botvinick, M. (2017). Neuroscience-Inspired Artificial Intelligence. *Neuron*, *95*(2), 245–258. https://doi.org/10.1016/j.neuron.2017.06.011
Hasson, U., Chen, J., & Honey, C. J. (2015). Hierarchical process memory: Memory as an integral component of information processing. *Trends in Cognitive Sciences*, *19*(6), 304–313. https://doi.org/10.1016/j.tics.2015.04.006
Hebblewhite, A., Hohwy, J., & Drummond, T. (2021). Events and Machine Learning. *Topics in Cognitive Science*, *13*(1), 243–247. https://doi.org/10.1111/tops.12520
Helmholtz H. (1867). Handbuch der Physiologischen Optik (English tranl.: 1924 JPC 583 Southall as Treatise on Physiological Optics) Voss.
Henke, K. (2010). A model for memory systems based on processing modes rather than consciousness. *Nature Reviews Neuroscience*, *11*(7), 523–532. https://doi.org/10.1038/nrn2850
Herszage, J., & Censor, N. (2017). Memory Reactivation Enables Long-Term Prevention of Interference. *Current Biology*, *27*(10), 1529-1534.e2. https://doi.org/10.1016/j.cub.2017.04.025
Herszage, J., & Censor, N. (2018). Modulation of Learning and Memory: A Shared Framework for Interference and Generalization. *Neuroscience*, *392*, 270–280. https://doi.org/10.1016/j.neuroscience.2018.08.006
Hertwig, R., & Engel, C. (2016). Homo Ignorans: Deliberately Choosing Not to Know. *Perspectives on Psychological Science*, *11*, 359–372. https://doi.org/10.1177/1745691616635594
Herzog, M. H., Drissi-Daoudi, L., & Doerig, A. (2020). All in Good Time: Long-Lasting Postdictive Effects Reveal Discrete Perception. *Trends in Cognitive Sciences*, *24*(10), 826–837. https://doi.org/10.1016/j.tics.2020.07.001
Hicks, J. M., & McDermott, J. H. (2024). Noise schemas aid hearing in noise. *Proceedings of the National Academy of Sciences*, *121*(47), e2408995121. https://doi.org/10.1073/pnas.2408995121
Hintzman, D. L. (2003). Robert Hooke's model of memory. *Psychonomic Bulletin & Review*, *10*(1), 3–14. https://doi.org/10.3758/BF03196465
Hintzman, D. L. (2016). Is memory organized by temporal contiguity? *Memory & Cognition*, *44*(3), 365–375. https://doi.org/10.3758/s13421-015-0573-8

Hochreiter, S & Schmidhuber, J. (1997). Long short-term memory. *Neural Computation*, *9*, 1735–1780.
Hollingworth, H.L. (1910). *The Journal of Philosophy, Psychology and Scientific Methods*, 7, 709-714.
Hommel, B. (2005). Perception in action: Multiple roles of sensory information in action control. *Cognitive Processing*, *6*(1), 3–14. https://doi.org/10.1007/s10339-004-0040-0
Hooke, Robert (1682). Lecture at the Royal Society of London, reviewed by Hintzman (2003).
Howard, M. W., & Kahana, M. J. (2002). A Distributed Representation of Temporal Context. *Journal of Mathematical Psychology*, *46*(3), 269–299. https://doi.org/10.1006/jmps.2001.1388
Howard, M. W. (2018). Memory as Perception of the Past: Compressed Time in Mind and Brain. *Trends in Cognitive Sciences*, *22*(2), 124–136. https://doi.org/10.1016/j.tics.2017.11.004
Hua, Y., Xiao, B., Veeravalli, B., & Feng, D. (2012). Locality-Sensitive Bloom Filter for Approximate Membership Query. *IEEE Transactions on Computers*, *61*(6), 817–830. https://doi.org/10.1109/TC.2011.108
Husserl, E (1962). The phenomenology of internal time consciousness, edited by M. Heidegger and translated by J. C. Churchill, Indiana University Press, https://doi.org/10.2307/j.ctvh4zhv9.
Irish, M., & Vatansever, D. (2020). Rethinking the episodic-semantic distinction from a gradient perspective. *Current Opinion in Behavioral Sciences*, *32*, 43–49. https://doi.org/10.1016/j.cobeha.2020.01.016
Ivancovsky, T., Baror, S., & Bar, M. (2024). A shared novelty-seeking basis for creativity and curiosity. *Behavioral and Brain Sciences*, *47*, e89. https://doi.org/10.1017/S0140525X23002807
Jacobson, H. (1951). Information and the Human Ear. *The Journal of the Acoustical Society of America*, *23*(4), 463–471. https://doi.org/10.1121/1.1906788
James, W. (1890). The principles of psychology New York: Holt.
Jenkins, R., Dowsett, A. J., & Burton, A. M. (2018). How many faces do people know? *Proceedings of the Royal Society B: Biological Sciences*, *285*(1888), 20181319. https://doi.org/10.1098/rspb.2018.1319
Jovani R, Fortuna MA (2007) The shape of the past in the World Wide Web: Scale-free patterns and dynamics. Physica A: Statistical Mechanics and its Applications 385:683–688.
Kahana MJ. (2020). Computational models of memory search. *Annual Review of Psychology*, *71*, 107–138.
Kahneman, D., & Miller, D. T. (1986). Norm theory: Comparing reality to its alternatives. *Psychological Review, 93*(2), 136–153. https://doi.org/10.1037/0033-295X.93.2.136
Karp, R. M. (1992). *On-line algorithms vs off-line algorithms: How much is it worth to know the future?* (TR 92-044). International Computer Science Institute.
Kastellakis, G., Tasciotti, S., Pandi, I., & Poirazi, P. (2023). The dendritic engram. *Frontiers in Behavioral Neuroscience*, *17*, 1212139. https://doi.org/10.3389/fnbeh.2023.1212139
Kemp, C., & Tenenbaum, J. B. (2008). The discovery of structural form. *Proceedings of the National Academy of Sciences*, *105*(31), 10687–10692. https://doi.org/10.1073/pnas.0802631105
Khayat, N., & Hochstein, S. (2019). Relating categorization to set summary statistics perception. *Attention, Perception, & Psychophysics*, *81*(8), 2850–2872. https://doi.org/10.3758/s13414-019-01792-7
Khvostov, V., Ásgeirsson, Á. G., & Kristjánsson, Á. (2025). Explicit access to detailed representations of feature distributions. *Psychonomic Bulletin & Review*, *32*(6), 2903–2912. https://doi.org/10.3758/s13423-025-02716-3
Kirchner, H., & Thorpe, S. J. (2006). Ultra-rapid object detection with saccadic eye movements: Visual processing speed revisited. *Vision Research*, *46*(11), 1762–1776. https://doi.org/10.1016/j.visres.2005.10.002
Kim, G., Lewis-Peacock, J. A., Norman, K. A., & Turk-Browne, N. B. (2014). Pruning of memories by context-based prediction error. *Proceedings of the National Academy of Sciences*, *111*(24), 8997–9002. https://doi.org/10.1073/pnas.1319438111
Klyubin, A. S., Polani, D., & Nehaniv, C. L. (2005). All Else Being Equal Be Empowered. In M. S. Capcarrère, A. A. Freitas, P. J. Bentley, C. G. Johnson, & J. Timmis (Eds.), *Advances in Artificial Life* (Vol. 3630, pp. 744–753). Springer Berlin Heidelberg. https://doi.org/10.1007/11553090_75

Knoblauch, A., Körner, E., Körner, U., & Sommer, F. T. (2014). Structural Synaptic Plasticity Has High Memory Capacity and Can Explain Graded Amnesia, Catastrophic Forgetting, and the Spacing Effect. *PLoS ONE*, *9*(5), e96485. https://doi.org/10.1371/journal.pone.0096485

Kondo, M. (2014) *The life-changing Magic of Tidying up: The Japanese Art of Decluttering and Organizing*. New York: Ten Speed Press; ISBN 978-1607747307.

Kroes, M. C. W., & Fernández, G. (2012). Dynamic neural systems enable adaptive, flexible memories. *Neuroscience & Biobehavioral Reviews*, *36*(7), 1646–1666. https://doi.org/10.1016/j.neubiorev.2012.02.014

Kubovy M (2015) The Deep Structure of Lives. Philosophia Scientae pp. 153–176.

Kumar, A. A. (2021). Semantic memory: A review of methods, models, and current challenges. *Psychonomic Bulletin & Review*, *28*(1), 40–80. https://doi.org/10.3758/s13423-020-01792-x

Kumaran, D., Hassabis, D., & McClelland, J. L. (2016). What Learning Systems do Intelligent Agents Need? Complementary Learning Systems Theory Updated. *Trends in Cognitive Sciences*, *20*(7), 512–534. https://doi.org/10.1016/j.tics.2016.05.004

Kumle, L., Nobre, A. C., & Draschkow, D. (2025a). Sensorimnemonic decisions: Choosing memories versus sensory information. *Trends in Cognitive Sciences*, *29*(4), 311–313. https://doi.org/10.1016/j.tics.2024.12.010

Kumle, L., Kovoor, J., Watt, R. L., Boettcher, S. E. P., Nobre, A. C., & Draschkow, D. (2025b). Long-term memory facilitates spontaneous memory usage through multiple pathways. *Current Biology*, *35*(5), 1171-1179.e5. https://doi.org/10.1016/j.cub.2025.01.045

Kurby, C. A., & Zacks, J. M. (2008). Segmentation in the perception and memory of events. *Trends in Cognitive Sciences*, *12*(2), 72–79. https://doi.org/10.1016/j.tics.2007.11.004

Lachman R, Naus MJ. The episodic/semantic continuum in an evolved machine. *Behavioral and Brain Sciences*. 1984;7(2):244-246. doi:10.1017/S0140525X00044484

Landauer, T.K. (1986). How Much Do People Remember! Some Estimates of the Quantity of Learned Information in Long-term Memory. *Cognitive Science*, *10*, 477–493.

Landauer, T. K., & Dumais, S. T. (1997). A Solution to Plato's Problem: The Latent Semantic Analysis Theory of Acquisition, Induction, and Representation of Knowledge. *Psychological Review*, *104*, 211–240.

Langille, J. J., & Brown, R. E. (2018). The Synaptic Theory of Memory: A Historical Survey and Reconciliation of Recent Opposition. *Frontiers in Systems Neuroscience*, *12*, 52. https://doi.org/10.3389/fnsys.2018.00052

Langille, J. J., & Gallistel, C. R. (2020). Locating the engram: Should we look for plastic synapses or information-storing molecules? *Neurobiology of Learning and Memory*, *169*, 107164. https://doi.org/10.1016/j.nlm.2020.107164

Lau, J. S.-H., & Brady, T. F. (2018). Ensemble statistics accessed through proxies: Range heuristic and dependence on low-level properties in variability discrimination. *Journal of Vision*, 18.

Laughlin, S. (2001). Energy as a constraint on the coding and processing of sensory information. *Current Opinion in Neurobiology*, *11*(4), 475–480. https://doi.org/10.1016/S0959-4388(00)00237-3

Lechner, HA, Squire, LR, & Byrne, JH. (1999). 100 Years of Consolidation—Remembering Müller and Pilzecker. *Learning and Memory*, *6*, 77–87.

Leib, A. Y., Kosovicheva, A., & Whitney, D. (2016). Fast ensemble representations for abstract visual impressions. *Nature Communications*, *7*(1), 13186. https://doi.org/10.1038/ncomms13186

Lengyel, M., & Dayan, P. (2007). Hippocampal Contributions to Control: The Third Way. *NIPS'07: Proceedings of the 21st International Conference on Neural Information Processing Systems*, 889–896.

Lewandowsky, S., Oberauer, K., & Brown, G. D. A. (2009). No temporal decay in verbal short-term memory. *Trends in Cognitive Sciences*, *13*(3), 120–126. https://doi.org/10.1016/j.tics.2008.12.003

Lewandowsky, S., & Oberauer, K. (2015). Rehearsal in serial recall: An unworkable solution to the nonexistent problem of decay. *Psychological Review*, *122*(4), 674–699. https://doi.org/10.1037/a0039684

Lieder, F., & Griffiths, T. L. (2020). Resource-rational analysis: Understanding human cognition as the optimal use of limited computational resources. *Behavioral and Brain Sciences*, *43*, e1. https://doi.org/10.1017/S0140525X1900061X

Lin, Q., Li, Z., Lafferty, J., & Yildirim, I. (2024). Images with harder-to-reconstruct visual representations leave stronger memory traces. *Nature Human Behaviour*, *8*(7), 1309–1320. https://doi.org/10.1038/s41562-024-01870-3

Lindsay, G. W. (2020). Attention in Psychology, Neuroscience, and Machine Learning. *Frontiers in Computational Neuroscience*, *14*, 29. https://doi.org/10.3389/fncom.2020.00029

Llewellyn, S. (2013). Such stuff as dreams are made on? Elaborative encoding, the ancient art of memory, and the hippocampus. *Behavioral and Brain Sciences*, *36*(6), 589–607. https://doi.org/10.1017/S0140525X12003135

Loftus, E. F., & Loftus, G. R. (1980). On the Permanence of Stored Information in the Human Brain. *American Psychologist*, *35*, 409–420.

Logie, M. R., & Donaldson, D. I. (2021). Do doorways really matter: Investigating memory benefits of event segmentation in a virtual learning environment. *Cognition*, *209*, 104578. https://doi.org/10.1016/j.cognition.2020.104578

Lorenzi, C., Soares, C., & Vonner, T. (2001). Second-order temporal modulation transfer functions. *The Journal of the Acoustical Society of America*, *110*(2), 1030–1038. https://doi.org/10.1121/1.1383295

Ma, W. J., Beck, J. M., Latham, P. E., & Pouget, A. (2006). Bayesian inference with probabilistic population codes. *Nature Neuroscience*, *9*(11), 1432–1438. https://doi.org/10.1038/nn1790

Mahr, J. B., & Csibra, G. (2018). Why do we remember? The communicative function of episodic memory. *Behavioral and Brain Sciences*, *41*. https://doi.org/10.1017/S0140525X17000012

Manning, L., Cassel, D., & Cassel, J.-C. (2013). St. Augustine's Reflections on Memory and Time and the Current Concept of Subjective Time in Mental Time Travel. *Behavioral Sciences*, *3*(2), 232–243. https://doi.org/10.3390/bs3020232

Marr, David. (1970). Simple memory: A theory for archicortex. *Philosophical Transactions of the Royal Society B: Biological Sciences*, *262*, 23–81.

Marr, D. (1982). Vision: A computational investigation into the human representation and processing of visual information. Cambridge, MA: MIT Press.

Mattar, M. G., & Lengyel, M. (2022). Planning in the brain. *Neuron*, *110*(6), 914–934. https://doi.org/10.1016/j.neuron.2021.12.018

Matthey, L., Bays, P. M., & Dayan, P. (2015). A Probabilistic Palimpsest Model of Visual Short-term Memory. *PLOS Computational Biology*, *11*(1), e1004003. https://doi.org/10.1371/journal.pcbi.1004003

McGaugh, J. L. (2015). Consolidating Memories. *Annual Review of Psychology*, *66*(1), 1–24. https://doi.org/10.1146/annurev-psych-010814-014954

McClelland, J. L., McNaughton, B.L, & O'Reilly, R. C. (1995). *Why There Are Complementary Learning Systems in the Hippocampus and Neocortex:InsightsFrom the Successesand Failuresof Connectionist Models of Learning and Memory*. *102*, 419–457.

McClelland, J. L., McNaughton, B. L., & Lampinen, A. K. (2020). Integration of new information in memory: New insights from a complementary learning systems perspective. *Philosophical Transactions of the Royal Society B: Biological Sciences*, *375*, 20190637.

McDermott JH, Schemitsch M, Simoncelli EP (2013) Summary statistics in auditory perception. Nature neuroscience 16:493–498.

McKenzie, S., & Eichenbaum, H. (2011). Consolidation and Reconsolidation: Two Lives of Memories? *Neuron*, *71*(2), 224–233. https://doi.org/10.1016/j.neuron.2011.06.037

McWalter R, Dau T (2017) Cascaded amplitude modulations in sound texture perception. Frontiers in Neuroscience 11:1–12.

McWalter R, McDermott JH (2018) Adaptive and Selective Time Averaging of Auditory Scenes. Current Biology 28:1405–1418.e10.

McWalter, R., & McDermott, J. H. (2019). Illusory sound texture reveals multi-second statistical completion in auditory scene analysis. *Nature Communications*, *10*(1), 5096. https://doi.org/10.1038/s41467-019-12893-0

Menon, V. (2023). 20 years of the default mode network: A review and synthesis. *Neuron*, *111*(16), 2469–2487. https://doi.org/10.1016/j.neuron.2023.04.023

Michael, E., De Gardelle, V., & Summerfield, C. (2014). Priming by the variability of visual information. *Proceedings of the National Academy of Sciences*, *111*(21), 7873–7878. https://doi.org/10.1073/pnas.1308674111
Mildner, J. N., & Tamir, D. I. (2019). Spontaneous Thought as an Unconstrained Memory Process. *Trends in Neurosciences*, *42*(11), 763–777. https://doi.org/10.1016/j.tins.2019.09.001
Molinaro, G., & Bar, M. (2025). Spontaneous thought as play: The value of fictional goals in the default mode network. *Current Opinion in Behavioral Sciences*, *63*, 101504. https://doi.org/10.1016/j.cobeha.2025.101504
Morton, N. W., Sherrill, K. R., & Preston, A. R. (2017). Memory integration constructs maps of space, time, and concepts. *Current Opinion in Behavioral Sciences*, *17*, 161–168. https://doi.org/10.1016/j.cobeha.2017.08.007
Moscovitch, M. (2008). The hippocampus as a 'stupid,' domain-specific module: Implications for theories of recent and remote memory, and of imagination. *Canadian Journal of Experimental Psychology/Revue Canadienne de Psychologie Expérimentale*, *62*(1), 62–79. https://doi.org/10.1037/1196-1961.62.1.62
Moscovitch, M., Cabeza, R., Winocur, G., & Nadel, L. (2016). Episodic Memory and Beyond: The Hippocampus and Neocortex in Transformation. *Annual Review of Psychology*, *67*(1), 105–134. https://doi.org/10.1146/annurev-psych-113011-143733
Mumford, D. (1992). On the computational architecture of the neocortex. *Biological Cybernetics*, *66*, 241–251.
Murty, V. P., DuBrow, S., & Davachi, L. (2019). Decision-making Increases Episodic Memory via Postencoding Consolidation. *Journal of Cognitive Neuroscience*, *31*(9), 1308–1317. https://doi.org/10.1162/jocn_a_01321
Nadel, L., Hupbach, A., Gomez, R., & Newman-Smith, K. (2012). Memory formation, consolidation and transformation. *Neuroscience & Biobehavioral Reviews*, *36*(7), 1640–1645. https://doi.org/10.1016/j.neubiorev.2012.03.001
Nagy, D. G., & Orban, G. (2016). Episodic memory as a prerequisite for online updates of model structure. *Proceedings of the 38th Annual Conference of the Cognitive Science Society*, 1–7.
Nagy, D. G., Török, B., & Orbán, G. (2020). Optimal forgetting: Semantic compression of episodic memories. *PLOS Computational Biology*, *16*(10), e1008367. https://doi.org/10.1371/journal.pcbi.1008367
Nairne, J. S., & Pandeirada, J. N. S. (2010). Adaptive memory: Ancestral priorities and the mnemonic value of survival processing. *Cognitive Psychology*, *61*(1), 1–22. https://doi.org/10.1016/j.cogpsych.2010.01.005
Nairne, J. S., Pandeirada, J. N. S., & Fernandes, N. L. (2017). Adaptive Memory. In *Learning and Memory: A Comprehensive Reference* (pp. 279–293). Elsevier. https://doi.org/10.1016/B978-0-12-809324-5.21060-2
Nelken, I., & De Cheveigné, A. (2013). An ear for statistics. *Nature Neuroscience*, *16*(4), 381–382. https://doi.org/10.1038/nn.3360
Nobre, A. C., & Stokes, M. G. (2019). Premembering Experience: A Hierarchy of Time-Scales for Proactive Attention. *Neuron*, *104*(1), 132–146. https://doi.org/10.1016/j.neuron.2019.08.030
Nobre, A. C., & Van Ede, F. (2023). Attention in flux. *Neuron*, *111*(7), 971–986. https://doi.org/10.1016/j.neuron.2023.02.032
Nørby, S. (2015). Why Forget? On the Adaptive Value of Memory Loss. *Perspectives on Psychological Science*, *10*(5), 551–578. https://doi.org/10.1177/1745691615596787
Norman-Haignere, S. V., Long, L. K., Devinsky, O., Doyle, W., Irobunda, I., Merricks, E. M., Feldstein, N. A., McKhann, G. M., Schevon, C. A., Flinker, A., & Mesgarani, N. (2022). Multiscale temporal integration organizes hierarchical computation in human auditory cortex. *Nature Human Behaviour*, *6*(3), 455–469. https://doi.org/10.1038/s41562-021-01261-y
Norris, D. (2017). Short-term memory and long-term memory are still different. *Psychological Bulletin*, *143*(9), 992–1009. https://doi.org/10.1037/bul0000108
Oberauer, K. (2019). Working Memory and Attention – A Conceptual Analysis and Review. *Journal of Cognition*, *2*(1), 36. https://doi.org/10.5334/joc.58
Olivers, C. N. L., & Roelfsema, P. R. (2020). Attention for action in visual working memory. *Cortex*, *131*, 179–194. https://doi.org/10.1016/j.cortex.2020.07.011

Orbán, G., Berkes, P., Fiser, J., & Lengyel, M. (2016). Neural Variability and Sampling-Based Probabilistic Representations in the Visual Cortex. *Neuron*, *92*(2), 530–543. https://doi.org/10.1016/j.neuron.2016.09.038
Orhan, A. E., & Jacobs, R. A. (2013). A Probabilistic Clustering Theory of the Organization of Visual Short-Term Memory. *Psychological Review*, *120*, 297–328.
Palmer, S. E., Marre, O., Berry, M. J., & Bialek, W. (2015). Predictive information in a sensory population. *Proceedings of the National Academy of Sciences*, *112*(22), 6908–6913. https://doi.org/10.1073/pnas.1506855112
Pan, R. K., Petersen, A. M., Pammolli, F., & Fortunato, S. (2018). The memory of science: Inflation, myopia, and the knowledge network. *Journal of Informetrics*, *12*(3), 656–678. https://doi.org/10.1016/j.joi.2018.06.005
Patihis, L., Frenda, S. J., LePort, A. K. R., Petersen, N., Nichols, R. M., Stark, C. E. L., McGaugh, J. L., & Loftus, E. F. (2013). False memories in highly superior autobiographical memory individuals. *Proceedings of the National Academy of Sciences*, *110*(52), 20947–20952. https://doi.org/10.1073/pnas.1314373110
Pitkow, X. (2016). Probability by Time. *Neuron*, *92*(2), 275–277. https://doi.org/10.1016/j.neuron.2016.10.007
Poirazi, P., & Mel, B. W. (2001). Impact of Active Dendrites and Structural Plasticity on the Memory Capacity of Neural Tissue. *Neuron*, *29*(3), 779–796. https://doi.org/10.1016/S0896-6273(01)00252-5
Polyn, S. M., & Cutler, R. A. (2017). Retrieved-context models of memory search and the neural representation of time. *Current Opinion in Behavioral Sciences*, *17*, 203–210. https://doi.org/10.1016/j.cobeha.2017.09.007
Potter, M. C. (1993). Very short-term conceptual memory. *Memory & Cognition*, *21*, 156–161.
Potter, M. C. (2012). Conceptual Short Term Memory in Perception and Thought. *Frontiers in Psychology*, *3*. https://doi.org/10.3389/fpsyg.2012.00113
Pouget, A., Beck, J. M., Ma, W. J., & Latham, P. E. (2013). Probabilistic brains: Knowns and unknowns. *Nature Neuroscience*, *16*(9), 1170–1178. https://doi.org/10.1038/nn.3495
Raaijmakers, J. G. W., & Shiffrin, R. M. (1981). Search of Associative Memory. *Psychological Review*, *88*, 93–134.
Radvansky, G. A., & Zacks, J. M. (2017). Event boundaries in memory and cognition. *Current Opinion in Behavioral Sciences*, *17*, 133–140. https://doi.org/10.1016/j.cobeha.2017.08.006
Rae, J. W., Potapenko, A., Jayakumar, S. M., & Lillicrap, T. P. (2019). *Compressive Transformers for Long-Range Sequence Modelling* (arXiv:1911.05507). arXiv. https://doi.org/10.48550/arXiv.1911.05507
Rae, J. W. (2021). *Towards Lifelong Reasoning with Sparse and Compressive Memory Systems*. University College London.
Raichle, M. E. (2010). Two views of brain function. *Trends in Cognitive Sciences*, *14*(4), 180–190. https://doi.org/10.1016/j.tics.2010.01.008
Raichle, M. E. (2015). The Brain's Default Mode Network. *Annual Review of Neuroscience*, *38*(1), 433–447. https://doi.org/10.1146/annurev-neuro-071013-014030
Raju, R. V., Guntupalli, J. S., Zhou, G., Wendelken, C., Lázaro-Gredilla, M., & George, D. (2024). Space is a latent sequence: A theory of the hippocampus. *Science Advances*, *10*(31), eadm8470. https://doi.org/10.1126/sciadv.adm8470
Raut, R. V., Snyder, A. Z., & Raichle, M. E. (2020). Hierarchical dynamics as a macroscopic organizing principle of the human brain. *Proceedings of the National Academy of Sciences*, *117*(34), 20890–20897. https://doi.org/10.1073/pnas.2003383117
Redondo, R. L., & Morris, R. G. M. (2011). Making memories last: The synaptic tagging and capture hypothesis. *Nature Reviews Neuroscience*, *12*(1), 17–30. https://doi.org/10.1038/nrn2963
Renoult, L., Irish, M., Moscovitch, M., & Rugg, M. D. (2019). From Knowing to Remembering: The Semantic–Episodic Distinction. *Trends in Cognitive Sciences*, *23*(12), 1041–1057. https://doi.org/10.1016/j.tics.2019.09.008
Renoult, L., & Rugg, M. D. (2020). An historical perspective on Endel Tulving's episodic-semantic distinction. *Neuropsychologia*, *139*, 107366. https://doi.org/10.1016/j.neuropsychologia.2020.107366

Rensink, R. A. (2014). Limits to the usability of iconic memory. *Frontiers in Psychology*, *5*. https://doi.org/10.3389/fpsyg.2014.00971
Richards, B. A., & Frankland, P. W. (2013). The conjunctive trace. *Hippocampus*, *23*(3), 207–212. https://doi.org/10.1002/hipo.22089
Richards, B. A., Lillicrap, T. P., Beaudoin, P., Bengio, Y., Bogacz, R., Christensen, A., Clopath, C., Costa, R. P., De Berker, A., Ganguli, S., Gillon, C. J., Hafner, D., Kepecs, A., Kriegeskorte, N., Latham, P., Lindsay, G. W., Miller, K. D., Naud, R., Pack, C. C., … Kording, K. P. (2019). A deep learning framework for neuroscience. *Nature Neuroscience*, *22*(11), 1761–1770. https://doi.org/10.1038/s41593-019-0520-2
Richards, B. A., & Frankland, P. W. (2017). The Persistence and Transience of Memory. *Neuron*, *94*(6), 1071–1084. https://doi.org/10.1016/j.neuron.2017.04.037
Robertson, E. M. (2009). From Creation to Consolidation: A Novel Framework for Memory Processing. *PLoS Biology*, *7*(1), e1000019. https://doi.org/10.1371/journal.pbio.1000019
Robertson, E. M. (2018). Memory instability as a gateway to generalization. *PLOS Biology*, *16*(3), e2004633. https://doi.org/10.1371/journal.pbio.2004633
Robinson, A. (2018). Einstein said that—Didn't he? *Nature*, *557*, 30.
Rodriguez-Ortiz, C. J., & Bermúdez-Rattoni, F. (2017). Determinants to trigger memory reconsolidation: The role of retrieval and updating information. *Neurobiology of Learning and Memory*, *142*, 4–12. https://doi.org/10.1016/j.nlm.2016.12.005
Roediger, H. L., & Uner, O. (2024). Critical Concepts in the Study of Learning and Memory. In Kahana MJ & Wagner AD, *Oxford handbook of human memory*. https://doi.org/10.1093/oxfordhb/9780190917982.013.1
Roüast, N. M., & Schönauer, M. (2023). Continuously changing memories: A framework for proactive and non-linear consolidation. *Trends in Neurosciences*, *46*(1), 8–19. https://doi.org/10.1016/j.tins.2022.10.013
Rouhani, N., Norman, K. A., Niv, Y., & Bornstein, A. M. (2020). Reward prediction errors create event boundaries in memory. *Cognition*, *203*, 104269. https://doi.org/10.1016/j.cognition.2020.104269
Rubin, D. C. (2006). The Basic-Systems Model of Episodic Memory. *Perspectives on Psychological Science*, *1*(4), 277–311. https://doi.org/10.1111/j.1745-6916.2006.00017.x
Rubin, D. C., & Umanath, S. (2015). Event memory: A theory of memory for laboratory, autobiographical, and fictional events. *Psychological Review*, *122*(1), 1–23. https://doi.org/10.1037/a0037907
Ryan, T. J., & Frankland, P. W. (2022). Forgetting as a form of adaptive engram cell plasticity. *Nature Reviews Neuroscience*, *23*(3), 173–186. https://doi.org/10.1038/s41583-021-00548-3
Sabat, M., Gouyette, H., Gaucher, Q., Lopez Espejo, M., David, S. V., Norman-Haignere, S., & Boubenec, Y. (2025). Neurons in auditory cortex integrate information within constrained temporal windows that are invariant to the stimulus context and information rate. *bioRxiv*, 2025.02.14.637944. https://doi.org/10.1101/2025.02.14.637944
Sadeh, T., & Pertzov, Y. (2020). Scale-invariant Characteristics of Forgetting: Toward a Unifying Account of Hippocampal Forgetting across Short and Long Timescales. *Journal of Cognitive Neuroscience*, *32*(3), 386–402. https://doi.org/10.1162/jocn_a_01491
Salge, C., Glackin, C., Polani, D. (2014). Empowerment–An Introduction. In: Prokopenko, M. (eds) Guided Self-Organization: Inception. Emergence, Complexity and Computation, vol 9. Springer, Berlin, Heidelberg. https://doi.org/10.1007/978-3-642-53734-9_4
Sanborn, A. N., & Chater, N. (2016). Bayesian Brains without Probabilities. *Trends in Cognitive Sciences*, *20*(12), 883–893. https://doi.org/10.1016/j.tics.2016.10.003
Sanborn, A. N. (2017). Types of approximation for probabilistic cognition: Sampling and variational. *Brain and Cognition*, *112*, 98–101. https://doi.org/10.1016/j.bandc.2015.06.008
Savin, C., Dayan, P., & Lengyel, M. (2011). Two is better than one: Distinct roles for familiarity and recollection in retrieving palimpsest memories. *Advances in Neural Information Processing Systems*, 9.
Schacter, D. L. (1999). The seven sins of memory—Insights From Psychology and Cognitive Neuroscience. *American Psychologist*, *54*, 182–203.

Schacter, D. L., & Addis, D. R. (2007). The cognitive neuroscience of constructive memory: Remembering the past and imagining the future. *Philosophical Transactions of the Royal Society B: Biological Sciences*, *362*(1481), 773–786. https://doi.org/10.1098/rstb.2007.2087
Schacter, D. L., Addis, D. R., Hassabis, D., Martin, V. C., Spreng, R. N., & Szpunar, K. K. (2012). The Future of Memory: Remembering, Imagining, and the Brain. *Neuron*, *76*(4), 677–694. https://doi.org/10.1016/j.neuron.2012.11.001
Scharnowski, F., Hermens, F., Kammer, T., Öğmen, H., & Herzog, M. H. (2007). Feature Fusion Reveals Slow and Fast Visual Memories. *Journal of Cognitive Neuroscience*, *19*(4), 632–641. https://doi.org/10.1162/jocn.2007.19.4.632
Schlesinger, H. J. (1970). The place of forgetting in memory functioning. *Journal of the American Psychoanalytic Association, 18*(2), 358–371. https://doi.org/10.1177/000306517001800206
Schmidhuber, J. (2009). Driven by Compression Progress: A Simple Principle Explains Essential Aspects of Subjective Beauty, Novelty, Surprise, Interestingness, Attention, Curiosity, Creativity, Art, Science, Music, Jokes. *arXiv:0812.4360 [Cs]*. http://arxiv.org/abs/0812.4360
Scofield, J. E., & Johnson, J. D. (2022). The diminishing precision of memory for time. *Psychonomic Bulletin & Review*, *29*(1), 212–219. https://doi.org/10.3758/s13423-021-01984-z
Sekeres, M. J., Winocur, G., & Moscovitch, M. (2018). The hippocampus and related neocortical structures in memory transformation. *Neuroscience Letters*, *680*, 39–53. https://doi.org/10.1016/j.neulet.2018.05.006
Shearer, C. M., Rawson, A. B., Barron, H. C., & O'Reilly, J. X. (2025). Memory reactivation during rest forms shortcuts in a cognitive map. *Scientific Reports*, *15*(1), 24724. https://doi.org/10.1038/s41598-025-06742-y
Sherman, B. E., & Turk-Browne, N. B. (2020). Statistical prediction of the future impairs episodic encoding of the present. *Proceedings of the National Academy of Sciences*, *117*(37), 22760–22770. https://doi.org/10.1073/pnas.2013291117
Sherman, B. E., Turk-Browne, N. B., & Goldfarb, E. V. (2024). Multiple Memory Subsystems: Reconsidering Memory in the Mind and Brain. *Perspectives on Psychological Science*, *19*(1), 103–125. https://doi.org/10.1177/17456916231179146
Shevlin, H. (2020). Current controversies in the cognitive science of short-term memory. *Current Opinion in Behavioral Sciences*, *32*, 148–154. https://doi.org/10.1016/j.cobeha.2020.02.005
Sims, C. R. (2016). Rate–distortion theory and human perception. *Cognition*, *152*, 181–198. https://doi.org/10.1016/j.cognition.2016.03.020
Singh, I., Tiganj, Z., & Howard, M. W. (2018). Is working memory stored along a logarithmic timeline? Converging evidence from neuroscience, behavior and models. *Neurobiology of Learning and Memory*, *153*, 104–110. https://doi.org/10.1016/j.nlm.2018.04.008
Slaney, M., & Casey, M. (2008). Locality-Sensitive Hashing for Finding Nearest Neighbors [Lecture Notes]. *IEEE Signal Processing Magazine*, *25*(2), 128–131. https://doi.org/10.1109/MSP.2007.914237
Sleator, D. D., & Tarjan, R. E. (1985). Self-adjusting binary search trees. *Journal of the ACM*, *32*(3), 652–686. https://doi.org/10.1145/3828.3835
Smith, S. M., & Vela, E. (2001). Environmental context-dependent memory: A review and meta-analysis. *Psychonomic Bulletin & Review*, *8*(2), 203–220. https://doi.org/10.3758/BF03196157
Sommer, T. (2016). The Emergence of Knowledge and How it Supports the Memory for Novel Related Information. *Cerebral Cortex*, bhw031. https://doi.org/10.1093/cercor/bhw031
Spens, E., & Burgess, N. (2024). A generative model of memory construction and consolidation. *Nature Human Behaviour*, *8*(3), 526–543. https://doi.org/10.1038/s41562-023-01799-z
Squire, L. R., Knowlton, B., & Musen, G. (1993). The Structure and Organization of Memory. *Annual Review of Psychology*, *44*, 453–495.
Squire, L. R., & Alvarez, P. (1995). Retrograde amnesia and memory consolidation: A neurobiological perspective. *Current Opinion in Neurobiology*, *5*(2), 169–177. https://doi.org/10.1016/0959-4388(95)80023-9
Standing, L. (1973). Learning 10000 pictures. *Quarterly Journal of Experimental Psychology*, *25*(2), 207–222. https://doi.org/10.1080/14640747308400340
Stewart, Neil, Chater, Nick, & Brown, Gordon D.A. (2006). Decision by sampling. *Cognitive Psychology*, *53*, 1–26.

Stern, Y., Katz, R., & Sadeh, T. (2020). Explicit Sequence Memory in Recall of Temporally-structured Episodes. *Scientific Reports*, *10*(1), 2666. https://doi.org/10.1038/s41598-020-59472-8
Steyvers, M., & Tenenbaum, J. B. (2005). The Large-Scale Structure of Semantic Networks: Statistical Analyses and a Model of Semantic Growth. *Cognitive Science*, *29*(1), 41–78. https://doi.org/10.1207/s15516709cog2901_3
Storm, B. C., & Patel, T. N. (2014). Forgetting as a consequence and enabler of creative thinking. *Journal of Experimental Psychology: Learning, Memory, and Cognition*, *40*(6), 1594–1609. https://doi.org/10.1037/xlm0000006
Storm, B. C., & Soares, J. S. (2024). Memory in the Digital Age. In Kahana MJ & Wagner AD, *Oxford handbook of human memory*. https://doi.org/10.1093/oxfordhb/9780190917982.001.0001
Suchow, J. W., Fougnie, D., Brady, T. F., & Alvarez, G. A. (2014). Terms of the debate on the format and structure of visual memory. *Attention, Perception, & Psychophysics*, *76*(7), 2071–2079. https://doi.org/10.3758/s13414-014-0690-7
Sun, P., Chubb, C., Wright, C. E., & Sperling, G. (2018). High-capacity preconscious processing in concurrent groupings of colored dots. *Proceedings of the National Academy of Sciences*, *115*(52). https://doi.org/10.1073/pnas.1814657115
Sun, J., Li, J., & Zhang, H. (2019). Human representation of multimodal distributions as clusters of samples. *PLOS Computational Biology*, *15*(5), e1007047. https://doi.org/10.1371/journal.pcbi.1007047
Székely, A., Török, B., Kiss, M., & Janacsek, K. (2024). Identifying Transfer Learning in the Reshaping of Inductive Biases. *OPEN MIND*. https://doi.org/10.1162/opmi_a_00158
Tarder-Stoll, H., Baldassano, C., & Aly, M. (2024). The brain hierarchically represents the past and future during multistep anticipation. *Nature Communications*, *15*(1), 9094. https://doi.org/10.1038/s41467-024-53293-3
Tiganj, Z., Gershman, S. J., Sederberg, P. B., & Howard, M. W. (2019). Estimating Scale-Invariant Future in Continuous Time. *Neural Computation*, *31*(4), 681–709. https://doi.org/10.1162/neco_a_01171
Tishby, N., Polani, D. (2011). Information Theory of Decisions and Actions. In: Cutsuridis, V., Hussain, A., Taylor, J. (eds) Perception-Action Cycle. Springer Series in Cognitive and Neural Systems. Springer, New York, NY. https://doi.org/10.1007/978-1-4419-1452-1_19
Tonegawa, S., Liu, X., Ramirez, S., & Redondo, R. (2015). Memory Engram Cells Have Come of Age. *Neuron*, *87*(5), 918–931. https://doi.org/10.1016/j.neuron.2015.08.002
Tsotsos, JK. (1990). Analyzing vision at a complexity level. *Behavioral and Brain Sciences*, *13*, 423–469.
Tsotsos, JK. (2022). When We Study the Ability to Attend, What Exactly Are We Trying to Understand? *Journal of Imaging*, *8*(8), 212. https://doi.org/10.3390/jimaging8080212
Tulving, E. (1985b.). How Many Memory Systems Are There? *American Psychologist*.
Tulving, E (1985a). Memory and Consciousness. *Canadian Psychology*, *26*, 1-12. https://doi.org/10.1037/h0080017
Tulving, E. (1972). Episodic and semantic memory. In Tulving, E & Donaldson, W, *Organization of memory* (pp. 381–402). Academic Press.
Tulving, E. (2001). Episodic memory and common sense: How far apart? *Philosophical Transactions of the Royal Society of London. Series B: Biological Sciences*, *356*(1413), 1505–1515. https://doi.org/10.1098/rstb.2001.0937
Turner, R., & Sahani, M. (2007). Modeling Natural Sounds with Modulation Cascade Processes. *Advances in Neural Information Processing Systems*, 8.
Umanath, S., & Coane, J. H. (2020). Face Validity of Remembering and Knowing: Empirical Consensus and Disagreement Between Participants and Researchers. *Perspectives on Psychological Science*, *15*(6), 1400–1422. https://doi.org/10.1177/1745691620917672
Underwood, B. J. (1969). Attributes of memory. *Psychological Review*, *76*(6), 559–573. https://doi.org/10.1037/h0028143
Utochkin, I. S., & Wolfe, J. M. (2018). Visual search for changes in scenes creates long-term, incidental memory traces. *Attention, Perception, & Psychophysics*, *80*(4), 829–843. https://doi.org/10.3758/s13414-018-1486-y

Vaswani, A., Shazeer, N., Parmar, N., Uszkoreit, J., Jones, L., Gomez, A. N., Kaiser, L., & Polosukhin, I. (2020). *Attention Is All You Need* (arXiv:1706.03762). arXiv. https://doi.org/10.48550/arXiv.1706.03762
Vitale, F., Janzen, I., & McGrenere, J. (2018). Hoarding and Minimalism: Tendencies in Digital Data Preservation. *Proceedings of the 2018 CHI Conference on Human Factors in Computing Systems - CHI '18*, 1–12. https://doi.org/10.1145/3173574.3174161
Vives, M.-L., De Bruin, D., Van Baar, J. M., FeldmanHall, O., & Bhandari, A. (2023). Uncertainty aversion predicts the neural expansion of semantic representations. *Nature Human Behaviour*, *7*(5), 765–775. https://doi.org/10.1038/s41562-023-01561-5
Vorster, A. P., & Born, J. (2015). Sleep and memory in mammals, birds and invertebrates. *Neuroscience & Biobehavioral Reviews*, *50*, 103–119. https://doi.org/10.1016/j.neubiorev.2014.09.020
Waskom, M. L., & Kiani, R. (2018). Decision Making through Integration of Sensory Evidence at Prolonged Timescales. *Current Biology*, *28*(23), 3850-3856.e9. https://doi.org/10.1016/j.cub.2018.10.021
Weinrich H (1997) *Lethe - the art and critique of forgetting* Cornell University Press [translation 2004].
Whitney D, Leib AY (2018) Ensemble Perception. Annu. Rev. Psychol 691225:1–12.
Wiskott, L., & Sejnowski, T. J. (2002). Slow Feature Analysis: Unsupervised Learning of Invariances. *Neural Computation*, *14*(4), 715–770. https://doi.org/10.1162/089976602317318938
Wixted, J. T. (2004). The Psychology and Neuroscience of Forgetting. *Annual Review of Psychology*, *55*(1), 235–269. https://doi.org/10.1146/annurev.psych.55.090902.141555
Wixted, J. T. (2005). A Theory About Why We Forget What We Once Knew. *Current Directions in Psychological Science*, *14*(1), 6–9. https://doi.org/10.1111/j.0963-7214.2005.00324.x.
Wolfe, J. M. (2012). Saved by a Log: How Do Humans Perform Hybrid Visual and Memory Search? *Psychological Science*, *23*(7), 698–703. https://doi.org/10.1177/0956797612443968
Wolfe, J. M. (2020). Visual Search: How Do We Find What We Are Looking For? *Annual Review of Vision Science*, *6*(1), 539–562. https://doi.org/10.1146/annurev-vision-091718-015048
Wolpert, D. H. (1996). The Lack of A Priori Distinctions Between Learning Algorithms. *Neural Computation*, *8*(7), 1341–1390. https://doi.org/10.1162/neco.1996.8.7.1341
Woodman, G. F., Carlisle, N. B., & Reinhart, R. M. G. (2013). Where do we store the memory representations that guide attention? *Journal of Vision*, *13*(3), 1–1. https://doi.org/10.1167/13.3.1
Yatziv, T., & Kessler, Y. (2018). A two-level hierarchical framework of visual short-term memory. *Journal of Vision*, *18*(9), 2. https://doi.org/10.1167/18.9.2
Yonelinas, A. P., Aly, M., Wang, W.-C., & Koen, J. D. (2010). Recollection and familiarity: Examining controversial assumptions and new directions. *Hippocampus*, *20*(11), 1178–1194. https://doi.org/10.1002/hipo.20864
Yuille, A., & Kersten, D. (2006). Vision as Bayesian inference: Analysis by synthesis? *Trends in Cognitive Sciences*, *10*(7), 301–308. https://doi.org/10.1016/j.tics.2006.05.002
Zacks, J. M., Speer, N. K., Swallow, K. M., Braver, T. S., & Reynolds, J. R. (2007). Event perception: A mind-brain perspective. *Psychological Bulletin*, *133*(2), 273–293. https://doi.org/10.1037/0033-2909.133.2.273
Zacks, J. M. (2020). Event Perception and Memory. *Annual Review of Psychology*, *71*, 165–191.
Zacks, O., Ginsburg, S., & Jablonka, E. (2022). The Futures of the Past The Evolution of Imaginative Animals. *Journal of Consciousness Studies*, *29*(3), 29–61. https://doi.org/10.53765/20512201.29.3.029.
Zhang, W.-H., Wu, S., Josić, K., & Doiron, B. (2023). Sampling-based Bayesian inference in recurrent circuits of stochastic spiking neurons. *Nature Communications*, *14*(1), 7074. https://doi.org/10.1038/s41467-023-41743-3
Zhang, W., & Luck, S. J. (2009). Sudden Death and Gradual Decay in Visual Working Memory. *Psychological Science*, *20*(4), 423–428. https://doi.org/10.1111/j.1467-9280.2009.02322.x